\documentclass[11pt]{article}
\usepackage{jheppub}

\usepackage{graphicx}

\usepackage{amsthm,amsmath,amssymb,bbm}
\usepackage{dsfont}  
\usepackage{enumitem} 

\makeatletter
\def\@fpheader{\relax}
\makeatother

\preprint{IGC-18/5-1\\[-2.5em]}

\newlength{\bracewidth}

\usepackage{amssymb}
\usepackage{dcolumn}
\usepackage{bm}
\usepackage{dsfont}
\usepackage{color}
\usepackage[normalem]{ulem}
\usepackage{subfig}
\usepackage{mathtools}
\usepackage{scalerel,stackengine}

\def\hmath$#1${\texorpdfstring{{\rmfamily\textit{#1}}}{#1}}

\newcommand{\be}{\begin{equation}}
\newcommand{\ee}{\end{equation}}
\newcommand{\bea}{\begin{eqnarray}}
\newcommand{\eea}{\end{eqnarray}}

\newcommand{\ket}[1]{| #1 \rangle}
\newcommand{\bra}[1]{\langle #1 |}
\newcommand{\scalar}[2]{\left\langle #1 | #2 \right\rangle}
\newcommand{\vect}[1]{\boldsymbol{#1}}
\newcommand{\mean}[1]{\left\langle #1 \right\rangle}

\title{\huge Gluing polyhedra with entanglement\\ in loop quantum gravity}

\author[a]{Bekir Bayta\c{s},}
\emailAdd{bub188@psu.edu}
\author[a]{Eugenio Bianchi,}
\emailAdd{ebianchi@gravity.psu.edu}
\author[b]{Nelson Yokomizo\,}
\emailAdd{yokomizo@fisica.ufmg.br}

\affiliation[a]{Institute for Gravitation and the Cosmos \& Physics Department,\\ Penn State, University Park, PA 16802, USA}
\affiliation[b]{Departamento de F\'isica - ICEx, Universidade Federal de Minas Gerais, \\
CP 702, 30161-970, Belo Horizonte, MG, Brazil}

\abstract{
In a spin-network basis state, nodes of the graph describe un-entangled quantum regions of space, quantum polyhedra. In this paper we show how entanglement between intertwiner degrees of freedom  enforces gluing conditions for neighboring quantum polyhedra. In particular we introduce Bell-network states, entangled states defined via squeezed vacuum techniques. We study correlations of quantum polyhedra in a dipole, a pentagram and a generic graph. We find that vector geometries, structures with neighboring polyhedra having adjacent faces glued back-to-back, arise from Bell-network states. We also discuss the relation to Regge geometries. The results presented show clearly the role that entanglement plays in the gluing of neighboring quantum regions of space.
}

\begin{document}
\maketitle
\flushbottom

\newpage

\section{Introduction}
\label{sec1:introduction}

\begin{figure}[t] 
\begin {center}
 \includegraphics[height=13.35em]{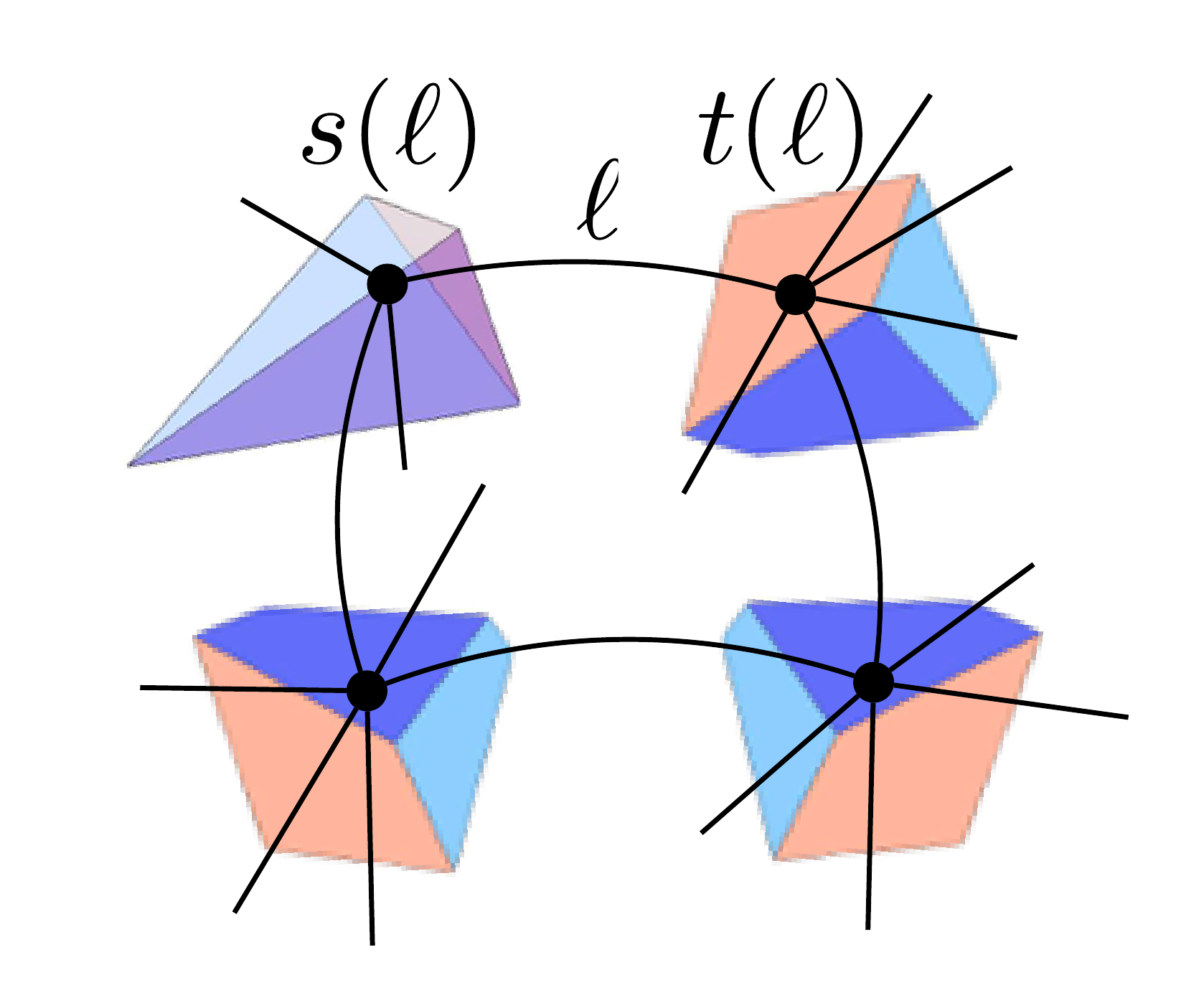} \hspace{2em}
 \includegraphics[height=13.35em]{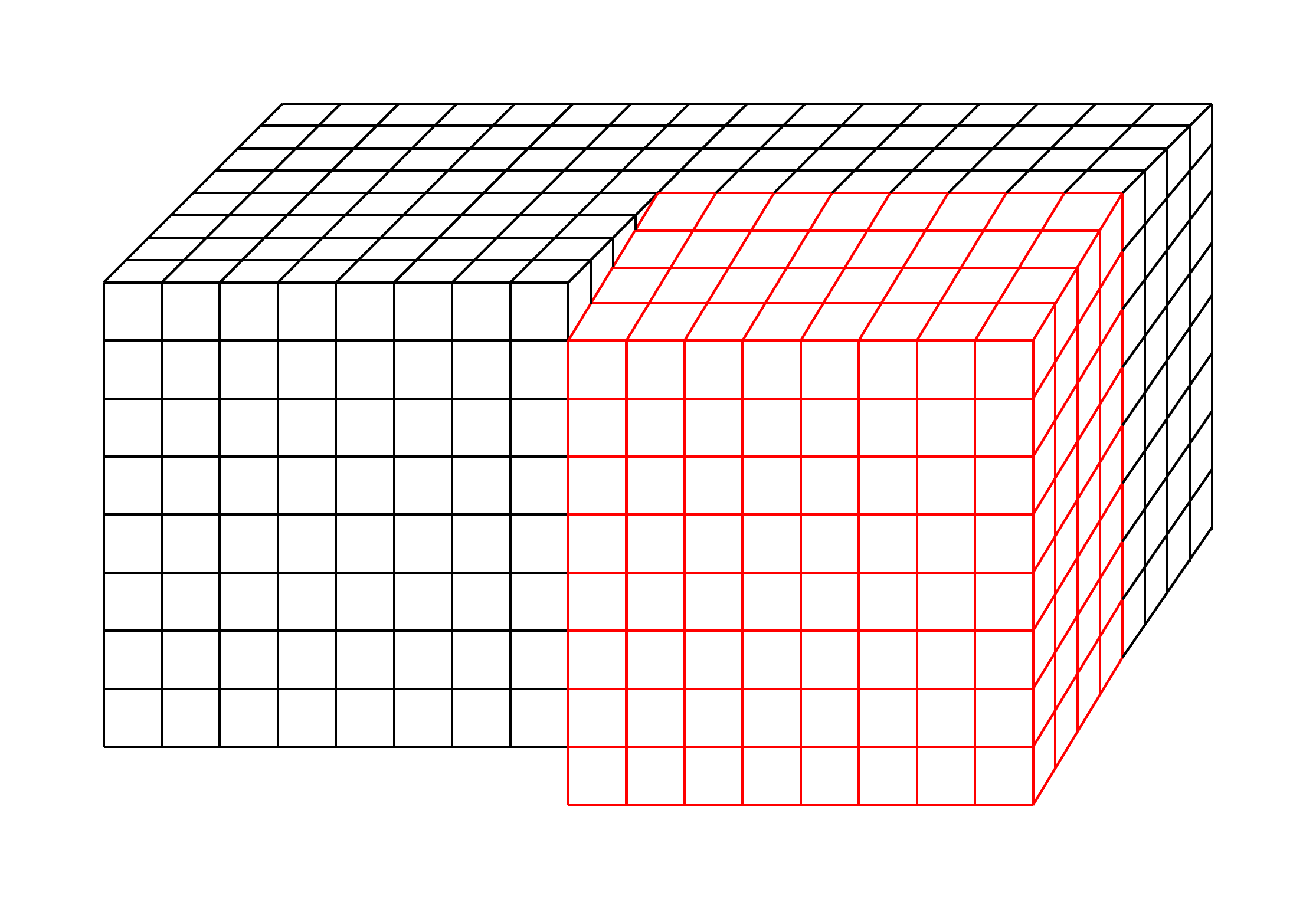}
\end{center}
\caption{\emph{(Left)} An example of a twisted geometry. Neighboring polyhedra have adjacent faces with the same area, but different shape; \emph{(Right)} An example of a vector geometry. Neighboring polyhedra have adjacent faces glued to each other: their normals are back-to-back.}
\label{fig:screw-dislocations}
\end{figure}

In loop quantum gravity, the geometry of space is quantized \cite{Rovelli:2004tv,Thiemann:2007zzarx,Ashtekar:2004eh}. Spin-network states provide an orthonormal basis of states for the quantum geometry of space. Nodes of the spin-network graph admit a geometric interpretation as \emph{quantum polyhedra} \cite{Bianchi:2010gc}. As a result, a discrete picture arises: a spin-network state can be understood as the quantum version of a collection of $3d$ Euclidean polyhedra. Each node of the spin-network graph corresponds to a polyhedron, and two polyhedra are said to be neighbors  if the two corresponding nodes are connected by a link (See Fig.~\ref{fig:screw-dislocations}). In this case, the source $s(\ell)$ and the target $t(\ell)$ of the link $\ell$ represent the two \emph{adjacent} faces of the two neighboring polyhedra. The classical degrees of freedom of the system are: 
\begin{itemize}
\item[i)] for each link $\ell$ of the graph, $(A_{\ell},\Theta_{\ell})$, the common area $A_\ell$ of the two adjacent faces and the extrinsic boost-angle $\Theta_{\ell}$ conjugated to this area;
\item[ii)] for each node $n$ of the graph, $(q_i,p_i)$, the $2F_n-6$ degrees of freedom that parametrize the phase space of a polyhedron with $F_n$ faces of fixed area. These degrees of freedom describe the shape of the polyhedron up to rescalings. For a given choice of frame, they encode the unit normals $\boldsymbol{n}$ to the faces of the polyhedron. 
\end{itemize}
The classical phase-space structure that arises from this construction is called a \emph{twisted geometry} \cite{Freidel:2010aq,Freidel:2010bw,Rovelli:2010km}. A typical point in phase space corresponds to a collection of largely uncorrelated polyhedra. Consider for instance two neighboring polyhedra: the shape of two adjacent faces will in general differ \cite{Dittrich:2008va,Bianchi:2008es}, while their area $A_\ell$ is constrained to be the same. The uncorrelated structure of the classical collection of polyhedra in a twisted geometry is reflected in the uncorrelated structure of a spin-network basis state in the quantum theory. A spin-network state $|\Gamma,i_n,j_l\rangle=\bigotimes_n |i_n\rangle$ is a tensor product of the intertwiner state $|i_n\rangle$ of each quantum polyhedron. In other words, quantum polyhedra in a spin-network state are \emph{un-entangled}.\\

This article focuses on configurations in phase space which have a geometric structure that is more rigid than the one of generic twisted-geometry configurations. In this family, the normals to the adjacent faces in neighboring polyhedra are back-to-back,
\begin{equation}
\vect{n}_{{s(\ell)}}=-\vect{n}_{t(\ell)}\,.
\label{eq:}
\end{equation}
When imposed consistently on all couples of neighboring polyhedra, this condition is non-trivial and defines a new structure called a $3d$ \emph{vector geometry} \cite{Barrett:2009as,Barrett:2009gg,Dona:2017dvf}. In a 3d vector geometry, the planes of the adjacent faces of neighboring polyhedra are consistently \emph{glued}, even though their shapes do not necessarily match. An example of a vector geometry is shown in Fig.~\ref{fig:screw-dislocations}. The condition of back-to-back normals is a constraint on the intrinsic shapes $(q_i,p_i)$ of the polyhedra, the degrees of freedom (ii) above. Note that no constraint on the extrinsic curvature $\Theta_{\ell}$ is imposed.\\

Vector geometries are related to the more familiar notion of Regge geometry \cite{Regge1961}. A 3d polyhedral Regge geometry \cite{Bianchi:2010gc} is obtained by imposing on a vector geometry the extra requirement that the shape of shared faces match, therefore defining an even more rigid structure. This hierarchy of 3d geometric structures is summarized in the table below:
\begin{table}[h!]
\begin{center}
\begin{tabular}{cll}
twisted geometry & = $\quad$ &phase space $\mathcal{M}_{\Gamma}$: area-matched polyhedra\\
$\cup$ & &\\
vector geometry & = &submanifold $\mathcal{V}_{\Gamma}\subset\mathcal{M}_{\Gamma}$: back-to-back normals\\
$\cup$ & &\\
polyhedral Regge geometry $\;$ & = &submanifold $\mathcal{R}_{\Gamma}\subset\mathcal{V}_{\Gamma}$: shape-matched polyhedra.\\[-1em]
\end{tabular}
\end{center}
\end{table}

Vector geometries arise in the study of semiclassical properties of spinfoam models \cite{Barrett:2009as,Barrett:2009gg,Dona:2017dvf,Han:2011re,Engle:2015zqa}. Our focus here is not the definition of a spinfoam vertex \cite{Engle:2007wy,Rovelli:2014ssa,Bianchi:2017hjl}, or a study of the dynamics of loop quantum gravity. Here we are interested in identifying states of the theory that describe the quantum geometry of 3d space --- both intrinsic and extrinsic --- and  reproduce the nearest-neighbor correlated structure of polyhedra in a classical vector geometry. We show that, in order to glue neighboring polyhedra, we have to entangle them. We introduce a class of states that represent quantum vector geometries and discuss their relation to Regge geometries.\\

Building a quantum version of a vector geometry requires entanglement. This is most easily explained in terms of a simple bipartite system consisting of two spin-$1/2$ particles, which we call the \emph{source} spin $s$ and the \emph{target} spin $t$ in analogy with the endpoints of a link in a spin-network graph. Let us consider the state
\begin{equation}
|p\rangle=|\!\uparrow\rangle_{s}\,|\!\downarrow\,\rangle_{t}\,,
\label{eq:}
\end{equation}
which is an eigenstate of the $z$-component of the spin. Clearly the expectation values of the spins are back-to-back on the state $|p\rangle$, i.e.
\begin{equation}
\langle p|\boldsymbol{J}_{s}| p\rangle\,=\,-\langle p|\boldsymbol{J}_{t}| p\rangle\,.
\label{eq:}
\end{equation}
However, the fluctuations of the two spins are uncorrelated and therefore, not back-to-back. This fact can be shown by taking into account the outcomes of a measurement. Suppose that we measure the $x$-component of the spin $s$ and find a positive value corresponding to the state $|\!\rightarrow\rangle_{s}$. The state of the spin $t$ after the measurement is still $|\!\downarrow\,\rangle_{t}$, which clearly is not back-to-back to $|\!\rightarrow\rangle_{s}$. 
This behavior is encoded in the spin correlation function
\begin{equation}
\textstyle	C^{\,ij}\equiv \langle  p| J^{i}_{s} J^{j}_{t}|p\rangle \,-\,\langle  p| J^{i}_{s}|p\rangle \langle p| J^{j}_{t}|p\rangle\;\;=\;0\,,
\label{eq:}
\end{equation}
which vanishes for all components $J^{i}$ of the spin. This is an immediate consequence of the fact that the state factorizes:  $|p\rangle $ is the product of a state for the subsystem $s$ and a state for the subsystem $t$.

To enforce the requirement that  spin fluctuations are back-to-back, we have to entangle the spins. Consider for instance the \emph{Bell state} $|\mathcal{B}\rangle$ \cite{opac-b1135621},
\begin{equation}
|\mathcal{B}\rangle =\frac{|\!\uparrow\rangle_{s}\,|\!\downarrow\,\rangle_{t}\,-\,|\!\downarrow\rangle_{s}\,|\!\uparrow\,\rangle_{t}}{\sqrt{2}}\,.
\label{eq:Bell}
\end{equation}
In this case, suppose that we measure the observable $\boldsymbol{n}\!\cdot\! \boldsymbol{J}_{s}$ of the spin $s$ and find a positive value corresponding to the eigenstate $|\!\nearrow\rangle_{s}$. The state of the spin $t$ after the measurement is now $|\!\swarrow\,\rangle_{t}$, which is back-to-back to the former. This happens for all directions $\boldsymbol{n}$ because the state $|\mathcal{B}\rangle$ is a singlet state: a state that satisfies
\begin{equation}
\big(\boldsymbol{J}_{s}+\boldsymbol{J}_{t}\big)^{2}\;|\mathcal{B}\rangle=0\,.
\label{eq:}
\end{equation}
The back-to-back behavior of spin fluctuations is encoded in the correlation function
\begin{equation}
\textstyle  C^{ij}\equiv \langle\mathcal{B} | J^{i}_{s} \,J^{j}_{t}|\mathcal{B}\rangle \;-\;\langle\mathcal{B}|J^{i}_{s}| B\rangle\,\langle\mathcal{B}|J^{j}_{t}|\mathcal{B}\rangle\;\;=\;-\frac{1}{4}\,\delta^{{ij}}\,,
\label{eq:}
\end{equation}
which is non-vanishing and negative, corresponding to the anti-correlation of fluctuations.\\

The correlations between the two subsystems can be quantified in information-theoretic terms using the entanglement entropy between subsystems. Given any two bounded observables $O_{s}$ and $O_{t}$ which probe only the subsystem $s$ or $t$, the rescaled correlation function
\begin{equation}
\mathcal{C}(O_{s},O_{t})=\frac{1}{\|O_{s}\| \|O_{t}\|}\Big(\langle \mathcal{B}|O_{s}\, O_{t}|\mathcal{B}\rangle-\langle \mathcal{B}|O_{s}|\mathcal{B}\rangle \langle \mathcal{B}|O_{t}|\mathcal{B}\rangle\Big)
\label{eq:}
\end{equation}
is bounded by the mutual information of the two subsystems \cite{wolf2008area},
\begin{equation}
\textstyle \frac{1}{2}\,\mathcal{C}(O_{s},O_{t})^{2}\;\leq\; S(\rho_{s})+S(\rho_{t})-S(\rho_{st})\,.
\label{eq:}
\end{equation}
Here $S(\rho)$ is the entanglement entropy of a subsystem with reduced density matrix $\rho$ and $S(\rho_{st}|\rho_{s}\!\otimes\! \rho_{t})=S(\rho_{s})+S(\rho_{t})-S(\rho_{st})$ is the mutual information between the subsystems $s$ and $t$. In the case of the product state $|p\rangle$, the mutual information vanishes and therefore the correlation functions of any two operators on $s$ and $t$ vanish. On the other hand, in the case of the Bell state $|\mathcal{B}\rangle$, the mutual information of $s$ and $t$ is non vanishing because of entanglement between the two and attains its maximum value $2\log 2$. The two spins in the Bell state $|\mathcal{B}\rangle$ are maximally entangled, a property which allows them to be always back-to-back.\\

Similarly to what happens for spins, gluing the adjacent faces of two neighboring quantum polyhedra requires entanglement. In this paper we use the formalism of squeezed spin-networks \cite{Bianchi:2016tmw,Bianchi:2016hmk} to build entangled states for neighboring quantum polyhedra. The idea can be illustrated by focusing on a single link of the spin-network graph. The bosonic Hilbert space of a link $\ell$ consists of four oscillators, two at the source and two at the target of the link \cite{Bianchi:2016hmk}. Denoting the creation operators $a^{\dagger}_{s}{}^{A}$ and $a^{\dagger}_{t}{}^{A}$, where $A=1,2$ is a spinor index, we define a Bell state of the link $\ell$ as
\begin{equation}
|\mathcal{B},\lambda\rangle_{\ell} =(1-|\lambda|^{2})\,\exp\big({\lambda\, \epsilon_{AB}a^{\dagger}_{s}{}^{A}a^{\dagger}_{t}{}^{B}}\big)\;|0\rangle_{s}|0\rangle_{t}\,,
\label{eq:}
\end{equation}
where $\lambda\in\mathbb{C}$ is a parameter that encodes the average area $A_{f}$ and the average extrinsic angle $\Theta_{f}$ of the link. A \emph{Bell spin-network} on a graph $\Gamma$ is defined as the gauge-invariant projection $P_{\Gamma}$ of a product of link states, i.e.,
\begin{equation}
|\Gamma,\mathcal{B},\lambda_{\ell}\rangle=P_{\Gamma}\;\bigotimes_{\ell\in \Gamma}|\mathcal{B},\lambda_{\ell}\rangle_{\ell}
\label{eq:}
\end{equation}
We investigate properties of Bell states for the dipole graph $\Gamma_{2}$, the pentagram graph $\Gamma_{5}$ and a general graph. We show that in the large spin limit, a Bell spin-network state represents a uniform superposition over classical vector geometries: a superposition over glued polyhedra. \\

Indications that entanglement in the degrees of freedom of the gravitational field play a crucial role for the emergence of a classical spacetime have surfaced in various approaches to nonperturbative quantum gravity \cite{VanRaamsdonk:2009ar,VanRaamsdonk:2010pw,Bianchi:2012ev,Jacobson:1995ab,Jacobson:2015hqa,Bianchi:2016tmw,Bianchi:2016hmk,Chirco:2017xjb,Livine:2017fgq}. The connectivity of space itself is argued to be related to the presence of entanglement among degrees of freedom in distinct regions of space  via holographic arguments \cite{VanRaamsdonk:2009ar,VanRaamsdonk:2010pw}. Quantum correlations also reflect metric properties of space in semiclassical gravity --- they provide its architecture --- as shown by the generic validity of an area law for the entanglement entropy of quantum fields in curved spaces, a property thus expected to hold for semiclassical states in any theory of quantum gravity \cite{Bianchi:2012ev}. Procedures for measuring distances and curvature from the network of quantum correlations have also been recently discussed in various emergent geometry scenarios \cite{Chirco:2017xjb,Saravani:2015moa,Cao:2016mst}. This paper explores quantum properties of the geometry of space and provides a concrete illustration of the relation between entanglement and geometry in loop quantum gravity.

\bigskip

The paper is organized as follows. In Section \ref{sec:classical-shape-matching} we discuss classical geometric structures on the phase space associated to a fixed graph. In Section \ref{sec:quantum-shape-matching}, we discuss Heisenberg uncertainty relations for quantum polyhedra and the uncorrelated structure of quantum twisted geometries. We introduce then a new class of states with nearest-neighbors entanglement --- Bell-network states. In Section \ref{sec:Gamma2} and \ref{sec:Gamma5}, we present a detailed analysis of how quantum polyhedra are glued in the entangled states on the simple graphs $\Gamma_{2}$ and $\Gamma_{5}$. We summarize our results and discuss generalizations in Section \ref{sec:summary}.

\section{Phase space and geometric structures on a graph}
\label{sec:classical-shape-matching}

The Hilbert space of loop quantum gravity (LQG) restricted to a graph $\Gamma$ can be understood as the quantization of a classical phase space with a finite number of degrees of freedom. In this section we discuss geometric structures in the graph phase space $\mathcal{M}_{\Gamma}$.

\subsection{The phase space of twisted geometries}

Consider a $3d$ manifold $\Sigma$, a cellular decomposition $\mathcal{C}(\Sigma)$ and its dual graph $\Gamma=\mathcal{C}(\Sigma)^{*}$ consisting of $N$ nodes and $L$ links. A simple example is given by a $3$-sphere decomposed in $5$ tedrahedral cells with dual graph $\Gamma_{5}=\mathcal{C}_{5}(\Sigma)^{*}$ given by the complete graph with $5$ nodes \cite{Rovelli:2014ssa}. When restricted to the graph $\Gamma$, the classical phase space $\mathcal{M}_\Gamma$ of loop quantum gravity is the direct product of link phase spaces, modulo gauge transformations at nodes $n$, 
\begin{equation}
\mathcal{M}_\Gamma=(\bigtimes_\ell \mathcal{M}_\ell)/\!\!/\vect{G}_{n}\,.
\label{eq:}
\end{equation} 
The phase space associated to a link $\ell$,
\begin{equation}
\mathcal{M}_\ell=T^*SU(2)\,,
\label{eq:}
\end{equation}
is a $SU(2)$ cotangent bundle associated with the $SU(2)$ configuration variable $g_\ell$ representing the holonomy of the Ashtekar connection $A_a^i$ along the link $\ell$ of the graph. The full classical phase space $\mathcal{M}$ of LQG on a smooth $3d$ manifold $\Sigma$ is the direct sum over graphs $\Gamma$ of the phase spaces $\mathcal{M}_\Gamma$. The restriction to a fixed graph corresponds to a truncation of the theory to a finite number of degrees of freedom \cite{Bianchi:2009tj} --- the holonomies along the links of $\Gamma$. Remarkably, despite the truncation, $\mathcal{M}_\Gamma$ still encodes a space of geometries, which are now discrete. They are known as twisted geometries and provide a generalization of the discrete geometries considered in Regge calculus \cite{Regge1961}. 

The interpretation of $\mathcal{M}_\Gamma$ in terms of twisted geometries relies on two ingredients. The first \cite{Freidel:2010aq,Freidel:2010bw} is the observation that the link phase space $ \mathcal{M}_\ell$ can be decomposed as
\begin{equation}
\mathcal{M}_\ell=S^2\times S^2\times T^*S^1
\label{eq:Mell}
\end{equation}
and parametrized in terms of phase-space variables
\begin{equation}
(\vect{n}_{s(\ell)}, \vect{n}_{t(\ell)},A_\ell,\Theta_\ell)\,,
\label{eq:}
\end{equation}
where $\vect{n}$ is a unit vector in $\mathbb{R}^{3}$. The second observation \cite{Bianchi:2010gc} is that a set of $F$ vectors that sums up to zero defines a Euclidean polyhedron with $F$ faces. Used together with the decomposition (\ref{eq:Mell}), this structure provides a decomposition of the LQG phase space in a Cartesian product
\begin{equation}
\mathcal{M}_{\Gamma} = (\bigtimes_\ell \mathcal{M}_\ell)/\!\!/\vect{G}_{n} = \bigtimes \limits_l T^*S^1 \, \bigtimes \limits_n \mathcal{S}_{F(n)}\,,
\label{eq:phasespace}
\end{equation}
where $\mathcal{S}_{F}$ is the phase space of a polyhedron with $F$ faces of fixed area. As a result, a configuration in the phase space $\mathcal{M}_{\Gamma}$  represents a twisted geometry --- a collection of $N$ polyhedra, one per node of the graph $\Gamma$.

In order to illustrate the degrees of freedom of a twisted geometry, it is useful to adopt the notation
\begin{equation}
\ell =(ab)\,,\quad s(\ell)=a\,,\quad t(\ell)=b\,,\quad a,b=1,\ldots,N\,.
\label{eq:}
\end{equation}
The degrees of freedom $(A_{ab},\Theta_{ab})$ represent the area $A_{ab}$ of the face $b$ of the polyhedron $a$, together with its conjugated momentum $\Theta_{ab}$. The condition $A_{ab}=A_{ba}$ reflects the fact that, in a twisted geometry, the area of the face $(ab)$ of neighboring polyhedra coincide. This is not the case for the shape of the face.

The shape of a face of a polyhedron is determined by a configuration in the phase space of $\mathcal{S}_{F(n)}$. This is the phase space of a convex Euclidean polyhedron with $F$ faces of fixed area $A_{ab}$. It can be parametrized in terms of \emph{flux} variables
\begin{equation}
\vect{E}_{ab}=A_{ab}\,\vect{n}_{ab}
\label{eq:}
\end{equation}
satisfying the closure constraint
\begin{equation}
\vect{G}_{a}=\sum_{b=1}^{F}\vect{E}_{ab}\,.
\label{eq:closure}
\end{equation}
As stated by the Minkowski theorem \cite{Minkowski}, a set of vectors  $\vect{E}_{ab}$ in
\be
\label{eq:Kapovich-Millson}
\mathcal{S}_{F(a)}= \big\{\vect{E}_{ab} \in S^2, b=1,\cdots,F \, \big| \, \vect{G}_{a} =0, \big\|\vect{E}_{ab} \big\|=A_{ab}\big\} \big/ SO(3)
\ee
identifies uniquely (up to rotations) a convex Euclidean polyhedron with $F$ faces of area $A_{ab}$ and unit normal $\vect{n}_{ab}$. The shape of the polyhedron can be reconstructed using the algorithm discussed in \cite{Bianchi:2010gc}. Moreover, $\mathcal{S}_F$ is naturally equipped with the structure of a phase space, known as the \textit{Kapovich-Millson} phase space \cite{Kapovich:1996}  where the rotationally-invariant Poisson brackets are obtained from functions of $\vect{E}_{ab}$ on $(S^2)^F$. Canonically conjugate variables 
\begin{equation}
(q_{ai},p_{aj})\,,\quad i,j=1,\ldots, 2(F-3)
\label{eq:qp}
\end{equation}
can be defined for instance by introducing the vector $\vect{p}_{ai}=\sum_{b=1}^{i+1} \vect{E}_{ab}$. Then we can define $q_{ai}$ as the angle between the vectors $\vect{p}_{ai} \times \vect{E}_{a\,i+1}$ and $\vect{p}_{ai} \times \vect{E}_{a\,i+2}$, and the conjugate momenta as the norms $p_{ai} = \|\vect{p}_{ai}\|$.

\subsection{Gluing polyhedra: vector geometries as a submanifold of $\mathcal{M}_{\Gamma}$} \label{sec:vec-geom}
In order to glue the faces of two polyhedra, we have to hold them so that they share a plane. In terms of the variables described above, the gluing condition is that the normals to the respective faces are back-to-back,
\begin{equation}
\vect{n}_{ab}=-\vect{n}_{ba}\,.
\label{eq:back-to-back}
\end{equation}
Note that the gluing condition does not require that the faces have the same shape: we can glue a tetrahedron to a cube. Even if the glued faces have the same shape, the gluing condition does not require that the edges of the two faces are aligned: two cubes can be glued with a twist. The gluing condition becomes non-trivial when, instead of having just two polyhedra, we have a collection of polyhedra that we want to glue.

\begin{figure}
\begin{center}
 \includegraphics[scale=0.4]{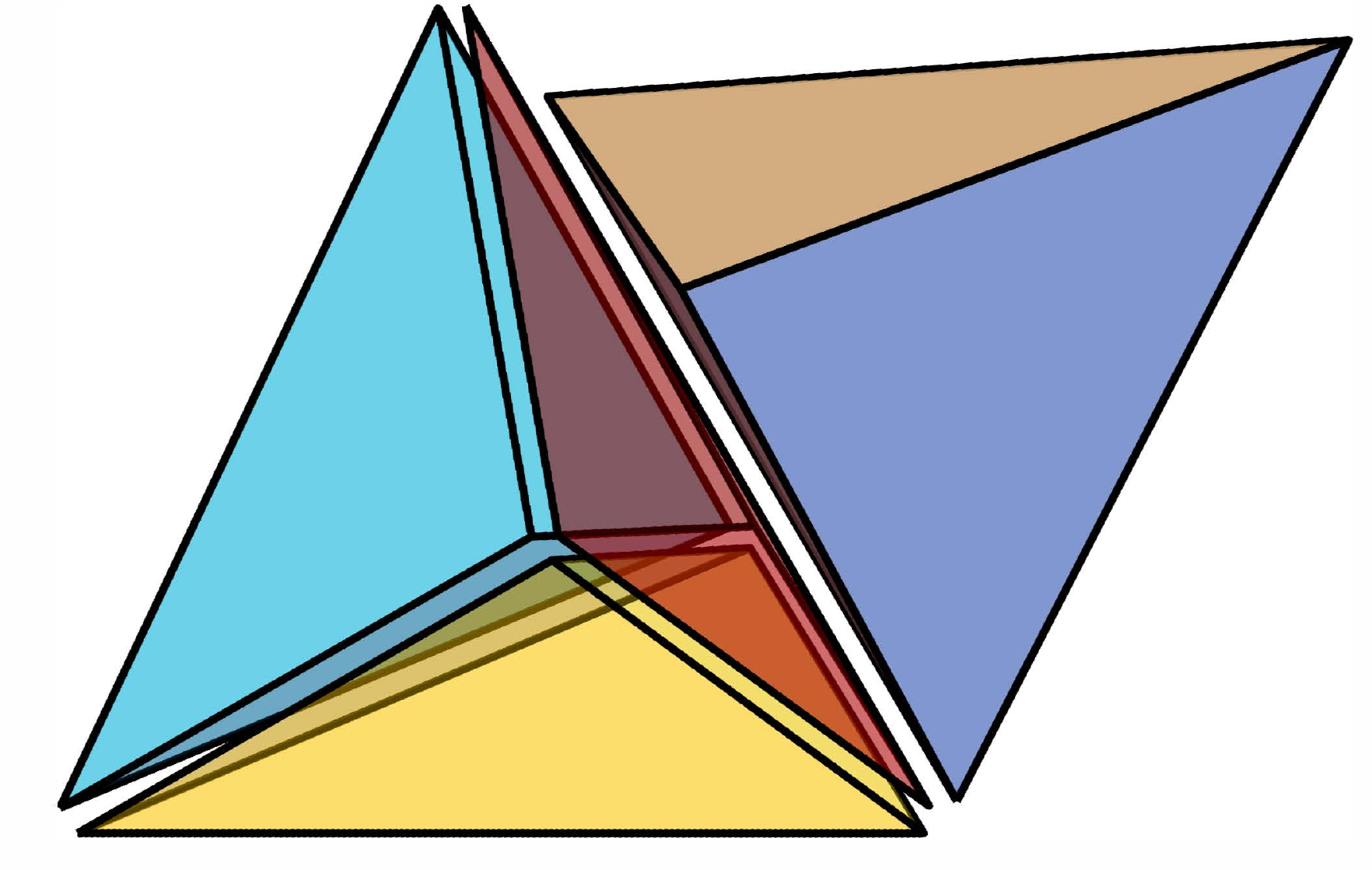}
\end{center} 
\caption{An example of vector geometry consisting of $5$ tetrahedra with adjacency relations encoded by the pentagram graph $\Gamma_{5}$. Adjacent faces have the same area and back-to-back normals. Note for instance that the downward-pointing face of the bottom tetrahedron and the upward-pointing face of the top tetrahedron have back-to-back normals. The set of back-to-back normals that describe the same vector geometry is shown in Fig.~\ref{fig:vector_geometry_graphical_v2}.}
\label{fig:vector-geometry}
\end{figure}

A twisted geometry consists of a collection of polyhedra with neighboring relations. Gluing neighboring polyhedra in a twisted geometry results in a geometric structure that is more rigid than the generic twisted geometry and is called a \emph{vector geometry} \cite{Barrett:2009as,Barrett:2009gg,Dona:2017dvf}.

Technically, a vector geometry is a twisted geometry $(A_{ab},\Theta_{ab}, \vect{n}_{ab}, \vect{n}_{ba})$ such that there exist $SO(3)$ elements $R_a$ at the nodes of $\Gamma$ that allow us to set
\be 
R_a \vect{n}_{ab} = - R_b \vect{n}_{ba} \, , \quad \forall \ell= (a,b) \, .
\label{eq:normal-matching}
\ee
The rotations $R_a$ can be used to fix a choice of local frame on each polyhedron. After acting with the $R_a$'s as gauge transformations, the normal-matching conditions reduce to the requirement that normals are back-to-back: the faces have parallel surfaces and can be glued together as in Fig.~\ref{fig:vector-geometry}. Accordingly, if all polyhedra in $\Gamma$ are isometrically embedded in $\mathbb{R}^3$, they can be rotated so that glued faces are always parallel with outwards pointing normals oriented in opposite directions. Note that this definition of a vector geometry is not formulated in terms of a constraint but in terms of an existence condition for the rotations $R_a$ in \eqref{eq:normal-matching}.

The conditions \eqref{eq:normal-matching} are defined in some chosen gauge to which the normals $(\vect{n}_{ab},\vect{n}_{ba})$ refer. Nonetheless, even though one cannot speak of normal vectors in the gauge-invariant phase space $\mathcal{M}_\Gamma$, there still is a clear notion of gauge-invariant vector geometry. The key point is that the condition \eqref{eq:normal-matching} defining vector geometries is gauge-invariant: if a set of normals $\{\vect{n}_{ab}\}$ forms a vector geometry, then its image $\{U_a \vect{n}_{ab}\}$ under gauge transformations $U_a \in SO(3)$ will also form a vector geometry satisfying \eqref{eq:normal-matching} for a new set of $R$'s. Hence, the space of vector geometries is naturally foliated as a union of gauge orbits. We denote the space of  gauge-invariant vector geometries by $\mathcal{V}_\Gamma$.

Vector geometries form a submanifold in the phase space of twisted geometries, $\mathcal{V}_\Gamma\subset \mathcal{M}_{\Gamma}$. We describe two procedures that provide a concrete description of this submanifold. The first procedure involves the choice of a trivial frame on a maximal tree of the graph $\Gamma$, as is often done in lattice gauge theory \cite{Montway:1997qf}. On the maximal tree, the back-to-back condition (\ref{eq:back-to-back}) can be imposed trivially. The normals associated to the leaves of the tree are now constrained: they are either be rotate to be back-to-back or not. We can now compute the gauge-invariant phase-space variables (\ref{eq:qp}) for a vector geometry \cite{Dona:2017dvf},
\begin{equation}
(q^{\mathrm{vec}}_{a\,i},p^{\mathrm{vec}}_{a\,i})\,,\qquad i=1,\ldots, 2(F_{a}-3),\;\;a=1,\ldots,N\,.
\label{eq:}
\end{equation}
This procedure provides a gauge-invariant characterization of a vector geometry in terms of the shapes of all the polyhedra present in the collection $\Gamma$. Fig.~\ref{fig:vector_geometry_graphical_v2} illustrates this procedure for the pentagram graph $\Gamma_{5}$ for which a vector geometry is shown.

The second procedure starts with the non-gauge invariant phase space $\bigtimes_\ell \mathcal{M}_\ell$. In this phase space, the gluing condition can be imposed as a constraint $\vect{T}_\ell$ for each link $\ell=(a,b)$,
\be
\label{eq:normal_constraint}
\vect{T}_\ell= \vect{n}_{ab} + \vect{n}_{ba} \approx  0 \, , \quad \forall \, l=(a,b) \, .
\ee
The solution for this set of constraints is a proper submanifold of the phase space $\bigtimes_\ell \mathcal{M}_\ell$. This is in fact the $4{L}$ dimensional Lagrangian submanifold  $\mathcal{A}_{\Gamma}=\bigtimes_\ell \mathcal{M}_\ell/\vect{T}_\ell$ studied in \cite{Aquilanti:2007:sw, Aquilanti:2012:sm, Haggard:2010:aw,Roberts:1998zka}. It is clear that any vector geometry satisfies the normal-matching constraints $\vect{T}_\ell$ in some gauge. Similarly, let $\mathcal{B}_{\Gamma} =(\bigtimes_\ell \mathcal{M}_\ell)/\vect{G}_n$ be the $(6L-3N)$-dimensional submanifold of $\mathcal{M}_\Gamma$ obtained by imposing the full set of closure constraints $\mathcal{G}_{n}$, without dividing by the gauge orbits. The submanifolds $\mathcal{A}_{\Gamma}$ and $\mathcal{B}_{\Gamma}$ are not phase spaces, since the constraint algebras do not close. The intersection $\mathcal{A}_{\Gamma} \cap \mathcal{B}_{\Gamma}$ describes simultaneous  solutions of both sets of constraints. In this submanifold, the set of back-to-back normals $(\vect{n}_{ab},-\vect{n}_{ab})$ at each link $l=(a,b)$ is selected so that the closure constraints hold at each node. We can now take gauge orbits of points in $\mathcal{A}_{\Gamma} \cap \mathcal{B}_{\Gamma}$. The space of such orbits is precisely the submanifold $\mathcal{V}_\Gamma$.

\subsection{Matching shapes: polyhedral Regge geometries as a submanifold of $\mathcal{V}_{\Gamma}$} 
Vector geometries can be seen as an assembly of polyhedra such that any pair of neighboring faces are glued back-to-back. Their shapes however can still be different. In order to obtain a continuous Regge geometry additional conditions must be imposed. We now turn to the description of the conditions that select the space $\mathcal{R}_{\Gamma}$ of polyhedral Regge geometries which is a submanifold of the space of vector geometries.

We say that two polyhedra are shape-matched if the glued faces are isometric polygons. A way to enforce this condition is to require for instance that the length of the edges and the planar angles between them in the two polygonal faces match. As edge-lengths and planar angles can be expressed in terms of the variables (\ref{eq:qp}), a polyhedral Regge geometry corresponds to a phase-space configuration
\begin{equation}
(q^{\mathrm{Regge}}_{ai},p^{\mathrm{Regge}}_{ai})\,,\qquad i=1,\ldots, 2(F_{a}-3),\;\;a=1,\ldots,N\,.
\label{eq:}
\end{equation}
An example of $3d$ Regge geometry is given by $5$ regular tetrahedra glued according to the relations encoded in the pentagram graph $\Gamma_{5}$. This geometry is parametrized by the shape $(q_{0},p_{0})$ of the regular tetrahedron \cite{Barbieri:1997ks,Baez:1999tk,Bianchi:2011ub,Bianchi:2012wb}, together with the area $A_{0}$ of its faces and the $10$ extrinsic angles $\Theta_{ab}$. This Regge geometry is a special case of a vector geometry as shown in Sec.~4 of \cite{Dona:2017dvf}.


\section{Gluing quantum polyhedra with entanglement}
\label{sec:quantum-shape-matching}
In LQG, the Hilbert space of states truncated to a fixed graph $\Gamma$ is $\mathcal{H}_{\Gamma}=L^{2}(SU(2)^{L}/SU(2)^{N})$. This space is spanned by spin-networks with graph $\Gamma$ and admits a decomposition in terms of spins and intertwiners,
\begin{equation}
\mathcal{H}_{\Gamma}=\bigoplus_{j_{\ell}}\Big(\bigotimes_{n}\mathcal{K}_n\Big)\,.
\label{eq:}
\end{equation}
This decomposition reflects the classical decomposition (\ref{eq:phasespace}) of the phase space $\mathcal{M}_{\Gamma}$ of twisted geometries on a graph. In particular, the $SU(2)$ intertwiner space $\mathcal{K}_n$ is the Hilbert space of a quantum polyhedron, the quantum version of the phase space (\ref{eq:Kapovich-Millson}). In this section we review the geometry of quantum polyhedra, show that in a spin-network basis state quantum shapes are uncorrelated, and introduce Bell-network states --- a family of states which describes glued quantum polyhedra and represents a quantum version of a vector geometry.

\subsection{Quantum polyhedra and the Heisenberg uncertainty relations}
\label{sec:quantum-tetrahedron}
Let us consider, within the graph $\Gamma$, a node $n$ of valency $F$. The intertwiner Hilbert space $\mathcal{K}_{n}$ is the invariant subspace of the tensor product of $F$ representation of $SU(2)$ associated to the links of $\Gamma$ at the node $n$,
\begin{equation}
\mathcal{K}_n(j_{n1},\ldots,j_{nF})=\mathrm{Inv}\big(\mathcal{H}^{(j_{n1})}\otimes\cdots\otimes \mathcal{H}^{(j_{nF})}\big)\,.
\label{eq:}
\end{equation}
The geometry of an intertwiner state $|i_{n}\rangle \in \mathcal{K}_{n}$ is determined by the flux operators
\begin{equation}
\vect{E}_{na}=\mathfrak{a}_{0}\,\vect{J}_{na}\,,
\label{eq:}
\end{equation}
defined in terms of $SU(2)$ generators $\vect{J}_{na}$ and the elementary area $\mathfrak{a}_{0}=8\pi G\hbar \gamma$ with Immirzi parameter $\gamma$. An intertwiner state $|i_{n}\rangle $ satisties
\begin{equation}
\vect{G}_{n} |i_{n}\rangle=0\,
\label{eq:}
\end{equation}
where $\vect{G}_{n}$ is the Gauss constraint
\begin{equation}
\vect{G}_{n}=\sum_{a=1}^{F} \vect{E}_{na}\,,
\label{eq:}
\end{equation}
the quantum version of the closure constraint (\ref{eq:closure}). The dimension of intertwiner space is
\begin{equation}
\dim \mathcal{K}_{n}\;=
\;\frac{1}{\pi}\int_0^{2\pi}\;\Bigg(\prod_{a=1}^F\frac{\sin\big((2j_{na}+1)\theta/2\big)}{\sin(\theta/2)}\Bigg)\; \sin^2(\theta/2)\;\,d\theta\,,
\label{eq:}
\end{equation}
and recoupling techniques provide an efficient way of building an orthonormal basis of $\mathcal{K}_{n}$.

States in $\mathcal{K}_{n}$ are quantum polyhedra \cite{Bianchi:2010gc} with $F$ faces of definite area: they are eigenstates of the area operator $A_{na}$,
\begin{equation}
A_{na}|i_{n}\rangle=\sqrt{\vect{E}_{na}\cdot \vect{E}_{na}}\,|i_{n}\rangle\,=\, \mathfrak{a}_{0}\,\sqrt{j_{na}(j_{na}+1)} |i_{n}\rangle\,.
\label{eq:}
\end{equation}
The quantum shape of the polyhedron is measured by the shape operator
\begin{equation}
g_{ab}(n)=\vect{E}_{na}\cdot \vect{E}_{nb}
\label{eq:}
\end{equation}
which in the Penrose spin-geometry theorem plays the role of a quantization of the metric \cite{spingeom,spingeom2,Bianchi:2010gc}. This operator measures the dihedral angle $\theta_{ab}(n)$ between the planes of the faces $(na)$ and $(nb)$ of the polyhedron \cite{Major:1999mc}.

Different components of the shape operator $g_{ab}(n)$ do not commute,
\be
[g_{ab}(n),g_{ac}(n)] = i \,\mathfrak{a}_{0} \;\vect{E}_{na} \!\cdot \!(\vect{E}_{nb} \times \vect{E}_{nc}) \, .
\ee
As a result of this non-commutativity, Heisenberg uncertainty relations for a quantum geometry follow: in any state $|i_{n}\rangle $, the dispersions $\Delta g_{ab}(n)$ in the quantum shape of the polyhedron satisfy the inequality
\be
\Delta g_{ab}(n) \, \Delta g_{ac}(n) \geq \frac{\mathfrak{a}_{0}}{2}  \, \bigg| \langle i_{n}| \vect{E}_{na} \cdot (\vect{E}_{nb} \times \vect{E}_{nc} ) |i_{n}\rangle \bigg| \, .
\label{eq:uncertainty-relations}
\ee
As a result, states with sharply defined features for the faces $(na)(nb)$, have maximal dispersion in the features of faces $(na)(nc)$ --- unless the three faces lie in a plane so that the right hand side of Eq.~(\ref{eq:uncertainty-relations}) vanishes. 

Coherent states for a quantum polyhedron can be built by starting with coherent spin states $|j,\vect{n}\rangle$, eigenstates of the spin $\vect{J}\cdot \vect{n}$ corresponding to the largest eigenvalue,  $\vect{J}\cdot \vect{n}|j,\vect{n}\rangle=+j|j,\vect{n}\rangle$, \cite{Livine:2007vk}. Choosing  a set of unit vectors $\vect{n}_{na}$ satisfying the closure condition $\sum_{a}j_{na} \vect{n}_{na}=0$ and projecting them to the gauge invariant subspace, one obtains the expression \cite{Livine:2007vk,Conrady:2009px}
\begin{equation}
|\Phi_{n}(\vect{n}_{na})\rangle=\int_{SU(2)}\!\!dg\;\;\bigotimes_{a=1}^{F}\Big(U(g)|j_{na},\vect{n}_{na}\rangle\Big)\,.
\label{eq:coherent-intertwiner}
\end{equation}
for a coherent intertwiner peaked on the shape of the classical polyhedron with normals $j_{na}\vect{n}_{na}$ \cite{Bianchi:2010gc}. Clearly, in a coherent state $|\Phi_{n}(\vect{n}_{na})\rangle$, fluctuations in the shape of the polyhedron are present as required by the uncertainty relations (\ref{eq:uncertainty-relations}).

\subsection{Quantum twisted geometries: spin-network basis states are un-entangled}
Spin-network basis states 
\begin{equation}
|\Gamma,j_{\ell},i_{n}\rangle =\bigotimes_{n}|i_{n}\rangle 
\label{eq:}
\end{equation}
provide an orthonormal basis of the graph Hilbert space $\mathcal{H}_{\Gamma}$. They are simultaneous eigenstates of the area operators and of a maximal commuting set of shape operators in the intertwiner space of each node. They represent quantum twisted geometries with definite area $A_{\ell}$, maximal dispersion of the extrinsic angle $\Theta_{\ell}$ and uncorrelated quantum shapes of polyhedra. 

Using coherent intertwiners $|\Phi_{n}(\vect{n}_{na})\rangle$, a semiclassical twisted geometry can be built: the spin-network state
\begin{equation}
|\Gamma,j_{\ell},\Phi_{n}(\vect{n}_{na})\rangle =\bigotimes_{n}|\Phi_{n}(\vect{n}_{na})\rangle 
\label{eq:GammaPhi}
\end{equation}
is peaked on a collection of polyhedra with average shape prescribed by the classical data encoded in the normals $\vect{n}_{ab}$. In particular the normals can be chosen so that a classical vector geometry is reproduced in average, or even a polyhedral Regge geometry. However, fluctuations around the average are uncorrelated. Suppose that we measure the shape of a polyhedron and find a given outcome. The shape of a neighboring polyhedron is uncorrelated, and therefore the two adjacent faces cannot be glued. This phenomenon can be made precise in terms of correlation functions. Let us consider operators $O_{n'}$ and $O_{n''}$ which measure the shape of the quantum polyhedra $n'$ and $n''$. The correlation function
\begin{equation}
\langle \Gamma,j_{\ell},\Phi_{n}| O_{n'}\;O_{n''}|\Gamma,j_{\ell},\Phi_{n}\rangle-\langle \Gamma,j_{\ell},\Phi_{n}| O_{n'}|\Gamma,j_{\ell},\Phi_{n}\rangle\langle \Gamma,j_{\ell},\Phi_{n}| O_{n''}|\Gamma,j_{\ell},\Phi_{n}\rangle\;=\;0
\label{eq:}
\end{equation}
vanishes despite the fact that the nodes $n'$ and $n''$ can be neighbors. Equivalently, for the state (\ref{eq:GammaPhi}), we can compute the mutual information of the nodes $n'$ and $n''$ and show that it vanishes. The geometry of quantum polyhedra in a spin-network state is un-entangled.\\

The bosonic representation of LQG \cite{Girelli:2005ii,Borja:2010rc,Livine:2011gp,Bianchi:2016tmw,Bianchi:2016hmk} provides a useful tool for illustrating the lack of rigidity of a quantum twisted geometry. In this representation, the Hilbert space $\mathcal{H}_{\Gamma}$ is obtained as a subspace of a bosonic Hilbert space $\mathcal{H}_{bos}$ describing $4L$ harmonic oscillators, where $L$ is the number of links in the graph. Explicitly, the bosonic Hilbert space is a tensor product of local Hilbert spaces attached to the endpoints of links:
\be
\mathcal{H}_{bos} = \bigotimes \limits_{\ell=1}^L \left(\mathcal{H}_{s(\ell)} \otimes \mathcal{H}_{t(\ell)} \right) \, ,
\ee
where each space $\mathcal{H}_{s(l)}$ and $\mathcal{H}_{t(\ell)}$ is associated with a pair of harmonic oscillators. As a result, there are four oscillators at each link, which we denote by $a^A_{s(\ell)}, a^B_{t(\ell)}$, $A, B = 0, 1$. We also use the notation $i=1,\ldots,2L$ to denote the seeds or endpoints of links. We then introduce link and node constraints:
\begin{align}
\mathcal{L}_{\ell} = \mathrm{I}_{s(\ell)} - \mathrm{I}_{t(\ell)} \approx 0\, , & \qquad \mathrm{I}_i = \frac{1}{2} \delta_{AB} \, a^{A \dagger}_i a^{B}_i \, , \\[1em]
\vect{G}_n = \sum \limits_{i \in n} \mathfrak{a}_{0}\vect{J}_i \approx 0 \, , & \qquad  \vect{J}_i = \frac{1}{2} \vect{\sigma}_{AB} \, a^{A \dagger}_i a^B_i \, . 
\label{eq:Jaa}
\end{align}
Bosonic states $\ket{s\,} \in \mathcal{H}_{bos}$ in general do not solve these constraints. The link constraint $\mathcal{L}_\ell$ matches the spins $j_{s(\ell)}=j_{t(\ell)}=j_\ell$ at the source and target of a link $l=(s,t)$, generating $U(1)$ transformations at each link. The node constraint $\vect{G}_n$ imposes invariance under $SU(2)$ gauge transformations at the node $n$. The LQG Hilbert space $\mathcal{H}_{\Gamma}$ is the proper subspace of $ \mathcal{H}_{bos}$ where these constraints are solved.

The vacuum state $|0\rangle_{\Gamma}$,
\begin{equation}
|0\rangle_{\Gamma}=\bigotimes_{i=1}^{2L}|0\rangle_{i}\,,\quad \text{with}\quad  a_{i}^{A}|0\rangle_{i}=0\,,
\label{eq:}
\end{equation}
satisfies all the constraints and is un-entangled as it is a product over the $2L$ seeds of the graph. A spin-network with coherent intertwiners is also un-entangled as it can be written as \cite{Bianchi:2016hmk}
\begin{equation}
 |\Gamma,j_\ell,\Phi_n\rangle=\sum_{m_i=-j_i}^{+j_i}\!\!\Big(\prod_n [\Phi_n]_{m_{(n,1)}\cdots m_{(n,F_{n})}}\Big)\Bigg(\prod_{i=1}^{2L} \frac{(a_i^{0\dagger})^{j_i-m_i}}{\sqrt{(j_i-m_i)!}}\frac{(a_i^{1\dagger})^{j_i+m_i}}{\sqrt{(j_i+m_i)!}} \Bigg) \; |0\rangle_{\Gamma} \, ,
\label{eq:spin-network-bosonic}
\end{equation}
which is a product over nodes of the graph. This formula shows again that a spin-network state with coherent intertwiners describes a quantum twisted geometry with no gluing of fluctuations of adjacent polyhedra.

\subsection{Entanglement and Bell-network states}
\label{sec:quantum-vector geometries}
Having clarified that, in order to glue quantum polyhedra we have to entangle them, we now move to the construction of a class of states with this property. \\

Squeezed vacua provide a powerful tool for capturing correlations in LQG \cite{Bianchi:2016hmk,Bianchi:2016tmw}. On a graph $\Gamma$ with $L$ links, a squeezed vacuum $\ket{\gamma} \in \mathcal{H}_{bos}$ is labeled by a squeezing matrix $\gamma_{AB}^{ij}$ which belongs to the Siegel disk $\mathcal{D}= \{ \gamma \in \mathrm{Mat}(4L,\mathbf{C}) | \gamma=\gamma^t \text{ and } \mathbbm{1}-\gamma \gamma^\dagger >0\}$ and encodes $2$-point correlation functions. The squeezed vacuum is defined by
\be
\ket{\gamma} = \mathrm{det}\big(\mathds{1} - \gamma \gamma^{\dagger}\big)^{1/4} \exp\left( \frac{1}{2} \gamma_{AB}^{i j} \, a_{i}^{A \dagger} a_{j}^{B \dagger}  \right) \ket{0} \, .
\ee
The indices $i,j=1,\dots,2L$ specify link endpoints, and $A,B=0,1$ distinguish between the two oscillators at a given link endpoint. Intuitively, a non-zero coefficient $\gamma_{ij}^{AB}$ of the squeezing matrix introduces correlations between the oscillator $A$ at $i$ and the oscillator $B$ at $j$. Note that the bosonic state $\ket{\gamma}$ is non-gauge-invariant and non-area-matched. A squeezed state in $\mathcal{H}_{\Gamma}$ is obtained by projection, $|\Gamma,\gamma\rangle =P_{\Gamma}|\gamma\rangle$. The projection can be implemented either via the use of the resolution of the identity in the spin-network basis,
\begin{equation}
P_{\Gamma}=\sum_{j_{\ell},i_{n}}|\Gamma,j_{\ell},i_{n}\rangle \langle \Gamma,j_{\ell},i_{n}|\,,
\label{eq:}
\end{equation}
or more directly via the loop expansion, $|\Gamma,\gamma\rangle =P_{\Gamma}|\gamma\rangle =\sum_{\Box}Z^{\phantom{\dagger}}_{\Box}\,F_{\Box}^{\dagger}|0\rangle_{\Gamma}$, as discussed in \cite{Bianchi:2016tmw}. Here we are interested in correlations between neighboring polyhedra, therefore we focus on link-wise squeezing. We consider a squeezing matrix  with a block-diagonal form with respect to the links of the graph, i.e., such that $\gamma_{ij}^{AB}=0$ for $i \neq j$.  The squeezing matrix is then given by
\be
\label{eq:squeezing_Bell_matrix}
\gamma_{AB}^{ij} = 
\begin{cases}
\lambda_\ell \, \epsilon_{AB}\;, 	&   \quad \text{if} \; i=t(\ell),j=s(\ell) \, ,  \\
- \lambda_\ell \, \epsilon_{AB}\;, 	&   \quad \text{if} \; i=s(\ell),j=t(\ell)  \, , \\
0 \;,			&   \quad \text{else} \, ,
\end{cases}
\ee
where $\lambda_\ell \in \mathbb{C}$, with $|\lambda_\ell|<1$. In the following we show that squeezed vacua with this structure describe glued polyhedra, a quantum version of the vector geometries discussed in Sec.~\ref{sec:vec-geom}. We call these states \emph{Bell-network} states.\\

A Bell-network state on a graph $\Gamma$ is parametrized by complex numbers $\lambda_{\ell}$ (one per link of the graph and with $|\lambda_{\ell}|<1$). It is given by
\begin{equation}
|\Gamma,\mathcal{B},\lambda_{\ell}\rangle =P_{\Gamma}\,\bigotimes_{\ell}|\mathcal{B},\lambda_{\ell}\rangle_{\ell}\,,
\label{eq:Bell-net1}
\end{equation}
where $|\mathcal{B},\lambda\rangle_{\ell}$ is the squeezed state 
\begin{equation}
|\mathcal{B},\lambda_{\ell}\rangle =(1-|\lambda_{\ell}|^{2})\,\exp\big({\lambda_{\ell}\, \epsilon_{AB}a^{\dagger}_{s}{}^{A}a^{\dagger}_{t}{}^{B}}\big)\;|0\rangle_{s}|0\rangle_{t}\,
\label{eq:Bell-net2}
\end{equation}
associated to a link of the graph. The geometric interpretation of the parameter $\lambda$ can be identified by computing the expectation values of the area operator and the holonomy operator on the link. We have that the expectation value of the area is
\begin{equation}
\langle \mathcal{B},\lambda_{\ell}|A_{\ell}|\mathcal{B},\lambda_{\ell}\rangle=\,\mathfrak{a}_{0}\,\sum_{j} \sqrt{j(j+1)} \;p_{j}(\lambda_{\ell})
\label{eq:}
\end{equation}
with $p_{j}(\lambda_{\ell}) = (1-|\lambda_{\ell}|^2)^2 \,(2 j +1) |\lambda_{\ell}|^{4j}$. In particular the expectation value diverges for $|\lambda_{\ell}|\to 1$ as it happens also for the expectation value of the spin
\begin{equation}
\langle \mathcal{B},\lambda_{\ell}|\,\mathrm{I}_{\ell}\,|\mathcal{B},\lambda_{\ell}\rangle=\sum_{j} j \;p_{j}(\lambda_{\ell})\;=\;\frac{|\lambda_\ell|^2}{1-|\lambda_\ell|^2} \, .
\label{eq:}
\end{equation}
The absolute value of $\lambda_\ell$ is thus fixed by the average spin at the link. In addition, the phase $\theta_\ell$ of $\lambda_\ell=|\lambda_\ell| e^{i \theta} $ is fixed by the mean value of the holonomy $h_\ell$ at $\ell$. In the bosonic representation, the holonomy operator is given by \cite{Bianchi:2016hmk,Livine:2011gp}:
\be \label{eq:hl}
(h_\ell)^A{}_B\equiv 
(2 \mathrm{I}_{t(\ell)}+1)^{-\frac{1}{2}}\big(
\epsilon_{AC}\, a_{t(\ell)}^{ C \dagger} \, a_{s(\ell)}^{ B \dagger}
-\epsilon_{BC}\, a_t^A\, a_s^C
\big)\,
(2 \mathrm{I}_{s(\ell)} +1)^{-\frac{1}{2}} \, .
\ee
For the Bell state $|\mathcal{B},\lambda_{\ell}\rangle$, we can compute the mean value of the trace of the holonomy $h_{\ell}$:
\be
\label{eq:mean-holonomy}
\langle \mathcal{B},\lambda_{\ell}|\,(h_\ell)^A{}_A\,|\mathcal{B},\lambda_{\ell}\rangle = - 2 \cos(\theta_\ell)\;  c(|\lambda_\ell|)\, , \qquad c(|\lambda_\ell|) = (1-|\lambda_\ell|^2)^2 \sum_{n=1}^\infty |\lambda_\ell|^{2n+1} \sqrt{n(n+1)} \, .
\ee
In the limit $|\lambda_\ell| \to 1$ of large spins, $c(|\lambda_\ell|)$ goes to $1$, and we have
\be
\label{eq:mean-holonomy2}
\langle \mathcal{B},\lambda_{\ell}|\,(h_\ell)^A{}_A\,|\mathcal{B},\lambda_{\ell}\rangle \simeq - 2 \cos (\theta_\ell) \, .
\ee
This approximation is quite accurate as soon as one leaves the Planck scale. As an illustration, for a mean spin of order $\mean{j_\ell} \simeq 10$, we already have $c(|\lambda_\ell|) \simeq 0.995$. We see that the phase of the squeezing parameter determines the mean value of the trace of the holonomy. \\

The state $|\mathcal{B},\lambda_{\ell}\rangle$ is a generalization of the Bell states (\ref{eq:Bell}) discussed in the introduction: it satisfies the condition
\begin{equation}
(\vect{J}_{s}+\vect{J}_{t})^{2}\;|\mathcal{B},\lambda_{\ell}\rangle=0\,,
\label{eq:}
\end{equation}
where $\vect{J}_{s}$ and $\vect{J}_{t}$ are defined in terms of bosonic operators in Eq.~(\ref{eq:Jaa}). Therefore, in a Bell state the fluxes at the source and at the target of a link are back-to-back, not only at the level of expectation values --- the fluctuations are anticorrelated, too. This is best shown by expanding the state over a spin basis,
\begin{equation}
|\mathcal{B},\lambda_{\ell}\rangle =(1- |\lambda_{\ell}|^2) \sum_{j}\sqrt{2j+1}\, \lambda_{\ell}^{2 j} \;|\mathcal{B},j\rangle  \,,
\label{eq:Bj}
\end{equation}
where $ |\mathcal{B},j\rangle$ is the maximally entangled state of spin $j$ (see App.~\ref{sec:mutual-info}),
\begin{equation}
 |\mathcal{B},j\rangle  =\frac{1}{\sqrt{2j+1}}\sum_{m=-j}^{+j} (-1)^{j-m}\; \ket{j, m}_{s} \ket{j, -m}_{t}\,,
\label{eq:}
\end{equation}
which has the same form as $(\ref{eq:Bell})$.

Using the decomposition (\ref{eq:Bj}), a Bell-network state can then be expressed as a sum over spins,
\begin{equation}
 |\Gamma,\mathcal{B},\lambda_{\ell}\rangle=\sum_{j_{\ell}} \left(\prod_{\ell} \big(1- |\lambda_{\ell}|^2\big) \sqrt{2j_{\ell}+1}\, \lambda_{\ell}^{2 j_{\ell}}\right)\; |\Gamma,\mathcal{B},j_{\ell}\rangle\,,
\label{eq:}
\end{equation}
where $|\Gamma, \mathcal{B},j_{\ell}\rangle$ has a remarkably simple form. It is a superposition of intertwiner states
\begin{equation}
|\Gamma,\mathcal{B},j_{\ell}\rangle=\sum_{i_{n}}\,\overline{A_{\Gamma}(j_{\ell},i_{n})}\;\bigotimes_{n}|i_{n}\rangle\,,
\label{eq:}
\end{equation}
with amplitude given by the \emph{symbol} of the graph $\Gamma$,
\be
\label{eq:projected-Bell-state-3}
A_{\Gamma}(j_{ab},i_a) = \sum \limits_{ \{m \}} \prod_{n} \big[i_n\big]^{m_{1} \cdots m_{F_{n}}} \, ,
\ee
i.e., the contraction of intertwining tensors $\big[i_n\big]^{m_{1} \cdots m_{F_{n}}}$ according to the combinatorics of the graph $\Gamma$. Techniques for computing the $SU(2)$ invariant amplitude $A_{\Gamma}$ for general graphs have been developed in \cite{Freidel:2012ji}, where a generating function was introduced in a coherent state representation. The asymptotic behaviour for large spins $j_{\ell}$ has also been recently investigated in \cite{Dona:2017dvf}, and such analysis can be applied to the study of Bell-network states in the limit of large average spin, $|\lambda_{\ell}|\to 1$.\\

The Bell-network state introduced here in Eq.~(\ref{eq:Bell-net1}) and (\ref{eq:Bell-net2}) provide a quantum version of the vector geometries discussed in Sec.~\ref{sec:vec-geom}: they are defined starting with objects that have back-to-back fluxes before projection. Clearly, to discuss the quantum gluing of polyhedra, one has to work at the gauge-invariant level, i.e., after projection.  In the next two sections we show the quantum gluing on specific examples for the graphs $\Gamma_{2}$ and $\Gamma_{5}$.

\section{Bell-network states: gluing quantum polyhedra on $\Gamma_{2}$}
\label{sec:Gamma2}
We describe the geometry of Bell-network states on the dipole graph.

\subsection{The dipole graph $\Gamma_{2}$}

\begin{figure}
\begin {center}
 \includegraphics[scale=1]{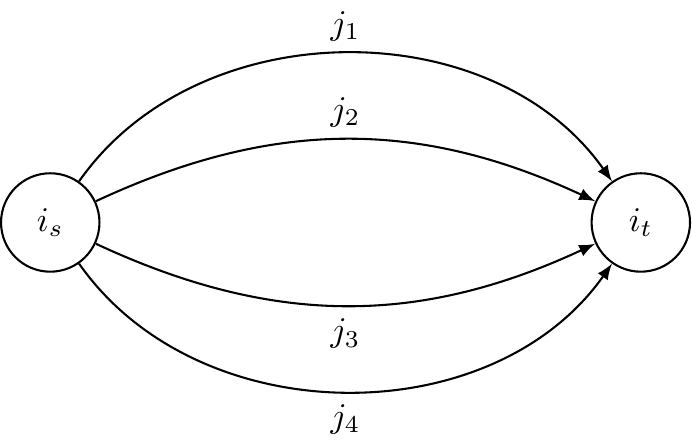}
\end{center}
\caption{Dipole graph $\Gamma_2$. The source ($s$) and target node ($t$)  are connected by four links $\ell=1,\dots,4$. The graph is dual to the triangulation of the 3-sphere formed by two tetrahedra glued along their boundaries. A spin-network state is labelled by spins $j_\ell$ and intertwiners $i_s,i_t$.}
\label{fig:dipole_graph}
\end{figure}

The dipole graph $\Gamma_2$ is formed by two nodes $n=s,t$ connected by four links $\ell=1,2,3,4$, as represented in Fig.~\ref{fig:dipole_graph}. The graph $\Gamma_2$ is dual to a triangulation of the three-sphere formed by two tetrahedra. The space of states of loop quantum gravity on $\Gamma_2$ is the Hilbert space $\mathcal{H}_{\Gamma_2} = L^2(SU(2)^4/SU(2)^2)$  of gauge-invariant $SU(2)$ states on the graph. An orthonormal basis for $\mathcal{H}_{\Gamma_2}$ is provided by spin-network states $\ket{i_s,i_t,j_\ell}$, where $j_\ell$ is the spin associated with the link $\ell$, the index $i_s$ labels an orthonormal basis of the intertwiner space $\mathcal{K}_{s}{(j_1,j_2,j_3,j_4)}$ associated with the node $s$ and similarly for the target node $t$. In the holonomy representation, a spin-network state is given by the wavefunction
\be
\psi_{i_{s} i_{t}j_{\ell}}(h_{\ell})=\scalar{h_\ell}{i_s,i_t,j_\ell} = \sum_{m_\ell,n_\ell} \prod_\ell \left[ D^{j_\ell}(h_\ell)\right]^{m_\ell}{}_{n_\ell} \left[i_s\right]^{n_1 n_2n_3 n_4} \left[i_t\right]_{m_1 m_2 m_3 m_4} \, , 
\ee
where spinor indices of intertwiners are lowered using the isomorphism $\epsilon^{(j)}:\mathcal{H}^{(j)} \to \mathcal{H}^{(j)}{}^*$ defined by $v_m = (-1)^{j-m} v^{-m}$. Intertwiners in this state are un-entangled: the state is factorized
\begin{equation}
|i_s,i_t,j_\ell\rangle =|i_{s}\rangle |i_{t}\rangle \,,
\label{eq:}
\end{equation}
and connected two-point functions of all gauge-invariant observables $g_{ab}(n)=\vect{E}_{na}\cdot \vect{E}_{nb}$ vanish,
\begin{align}
C_{ab\,cd}(n,n')\;=\;&\langle i_s,i_t,j_\ell|\, g_{ab}(n)\;g_{cd}(n')|i_s,i_t,j_\ell\rangle +\\[.5em]
&- \langle i_s,i_t,j_\ell|\, g_{ab}(n)|i_s,i_t,j_\ell\rangle\;\langle i_s,i_t,j_\ell|g_{cd}(n')|i_s,i_t,j_\ell\rangle\;=\;0\,.\nonumber
\label{eq:}
\end{align}
On the other hand, Bell-network states have non-trivial correlations as we now discuss.\\

To define Bell-network states, it is useful to introduce the bosonic representation of $\mathcal{H}_{\Gamma_2}$ \cite{Girelli:2005ii,Bianchi:2016hmk}, where each link is associated with four harmonic oscillators. We denote the Hilbert space of states of this collection of sixteen oscillators by $\mathcal{H}^{(16)}_{bos}$ and label each link endpoint by an index $i$. The space of states on the dipole graph $\Gamma_{2}$ is embedded unitarily in the oscillator model under the map:
\be
\sqrt{2j_\ell + 1} \left[ D^{j_\ell}(h_\ell)\right]^{m_\ell}{}_{n_\ell} \mapsto  (-1)^{j_\ell - n_\ell} \ket{j_\ell,m_\ell}_{t(\ell)} \ket{j_\ell,-n_\ell}_{s(\ell)}\, ,
\label{eq:schwinger-map-1}
\ee
where 
\be
\ket{j,m}_i = \frac{\left(a_i^{0 \dagger} \right)^{j + m} \left( a_i^{1 \dagger} \right)^{j - m} }{\sqrt{(j+m)! (j-m)!}} \ket{0}_i\, ,
\label{eq:schwinger-map-2}
\ee
and $\ket{0}_i$ is the local vacuum state annihilated by the operators $a_i^A$. The states $\ket{j,m}$ at each link endpoint are spin states with spin $j$ and magnetic number $m$, and \eqref{eq:schwinger-map-2} is the usual Schwinger oscillator model of angular momentum. Note that the map \eqref{eq:schwinger-map-1} is not surjective: the space of states of loop quantum gravity is a proper subspace $\mathcal{H}_{\Gamma_2} \subset \mathcal{H}^{(16)}_{bos}$ of the bosonic representation selected by the area-matching and Gauss constraints (see \cite{Bianchi:2016hmk}). The area-matching constraint is the requirement that only products of source and target local states $\ket{j_\ell,m_\ell}_t$ and $\ket{j_\ell',n_\ell}_s$ with the same spins are allowed, $j_\ell=j_\ell'$, as required by the map \eqref{eq:schwinger-map-1}. The Gauss constraint imposes gauge-invariance at each node $n$. We denote  the projection to the space of area-matched, gauge-invariant states by $P_{\Gamma_2}:\mathcal{H}^{(16)}_{bos} \to \mathcal{H}_{\Gamma_2}$.

\subsection{Bell-network state on $\Gamma_2$}

Consider the Bell-network state $\ket{\Gamma_{2},\mathcal{B},\lambda_{\ell}} \in \mathcal{H}^{(16)}_{bos}$ obtained by squeezing the vacuum state with the squeezing matrix $\lambda_\ell\, \epsilon_{AB}$ at each link, Eq.~\eqref{eq:Bell-net2}, and then projecting to the gauge-invariant subspace:
\begin{equation}
\ket{\Gamma_{2},\mathcal{B},\lambda_{\ell}}=\frac{1}{\mathcal{N}}\,P_{\Gamma_{2}}\,\exp\bigg(\sum_{\ell}{\lambda_{\ell}\, \epsilon_{AB}a^{\dagger A}_{s(l)}a^{\dagger B}_{t(\ell)}}\bigg)\;|0\rangle_{\Gamma_{2}}\,
\label{eq:}
\end{equation}
where $\mathcal{N}$ is a normalization. As the squeezing produces entangled bosonic pairs at the source and the target of a link, the squeezed vacuum is already area-matched. As a result, the projection operator $P_{\Gamma}$ acts non-trivially at nodes only and is given by $P_{\Gamma}=P_{t}\,P_{s}$, with
 \be
P_s = \sum_k \ket{i_k}\bra{i_k} \, ,
\label{eq:projection-node}
\ee
and $\ket{i_k} \in \mathcal{K}({j_1,j_2, j_3, j_4})$ an orthonormal basis of intertwiners at $s$. The Bell-network state on $\Gamma_{2}$ can then be expressed as
\begin{equation}
\ket{\Gamma_{2},\mathcal{B},\lambda_{\ell}}=\frac{1}{\mathcal{N}}\,\sum_{j_{\ell}}\big( \prod_\ell \lambda^{2 j_\ell} \big)\sqrt{\dim\mathcal{K}(j_{\ell})}\;\ket{\Gamma_{2},\mathcal{B},j_{\ell}}
\label{eq:}
\end{equation}
where $\dim\mathcal{K}(j_{\ell})$ is the dimension of the $4$-valent intertwiner space,
\begin{equation}
\dim\mathcal{K}(j_{\ell})=\min\big(j_{1}+j_{2}-|j_{1}-j_{2}|\,,\,j_{3}+j_{4}-|j_{3}-j_{4}|\big)\,,
\label{eq:}
\end{equation}
and $\ket{\Gamma_{2},\mathcal{B},j_{\ell}}$ is the Bell-network state at fixed spins,
\be
\ket{\Gamma_2, \mathcal{B},j_\ell} = \frac{1}{\sqrt{\dim \mathcal{K}(j_{\ell})}} \sum_{k=1}^{\dim \mathcal{K}(j_{\ell})} \ket{i_k}_t \ket{\tilde{i}_k}_s \, .
\label{eq:gamma-fixed-j}
\ee
Note that the intertwiner $\tilde{i}_k$ is obtained from $\ket{i_k}$ by acting on all intertwiner indices with the antilinear map $\zeta:\mathcal{H}^{(j)} \to \mathcal{H}^{(j)}$ defined by
\be
\tilde{v}^m = (\zeta v)^m = (v^{-m})^* (-1)^{j-m} \, .
\label{eq:time-reversal}
\ee
This map \eqref{eq:time-reversal} corresponds to the operation of time-reversal for spin states in $\mathcal{H}^{(j)}$. Hence, if the states $\ket{i_k}$ form an orthonormal set of eigenstates of an observable $\hat{O}$, then the states $\ket{\tilde{i}_k}$ form an orthonormal basis of eigenstates of the time-reversed operator $\zeta \hat{O} \zeta^{-1}$. Now, the action of time-reversal on area operators $\vect{E}_{na}$ only amounts to a change of sign, $\zeta \vect{E}_{na} \zeta^{-1} = - \vect{E}_{na}$, and as a result the Penrose metric operator $g_{ab}$ is not affected by this operation. Since any observable $\hat{O}$ of the intrinsic geometry can be written in terms of components of the Penrose metric, it follows that the intertwiners $\ket{i_k}$ and $\ket{\tilde{i}_k}$ describe the same local intrinsic geometry. This leads to a simple geometric interpretation of the projected state $\ket{\Gamma_2, \mathcal{B},j_\ell}$: it consists of a perfectly correlated state such that if the measurement of an observable $\hat{O}$ of the geometry  of the quantum tetrahedron $s$ has an outcome $o_k$, then the observation of the same quantity at the quantum tetrahedron $t$ gives the same result. 

Note that this property is valid for any local observable $\hat{O}$, since any basis of orthonormal intertwiners can be used in Eq.~\eqref{eq:projection-node}. Such a behavior mirrors that of Bell states for a pair of spin $1/2$ systems, where observations of the individual spins $\vect{J}\cdot \vect{n}$ are perfectly correlated for measurements performed in arbitrary directions $\vect{n}$. In the present case, instead of spins, each subsystem is a quantum tetrahedron, and observations of individual shapes are perfectly correlated.

\begin{figure}
\begin {center}
 \includegraphics[scale=.3]{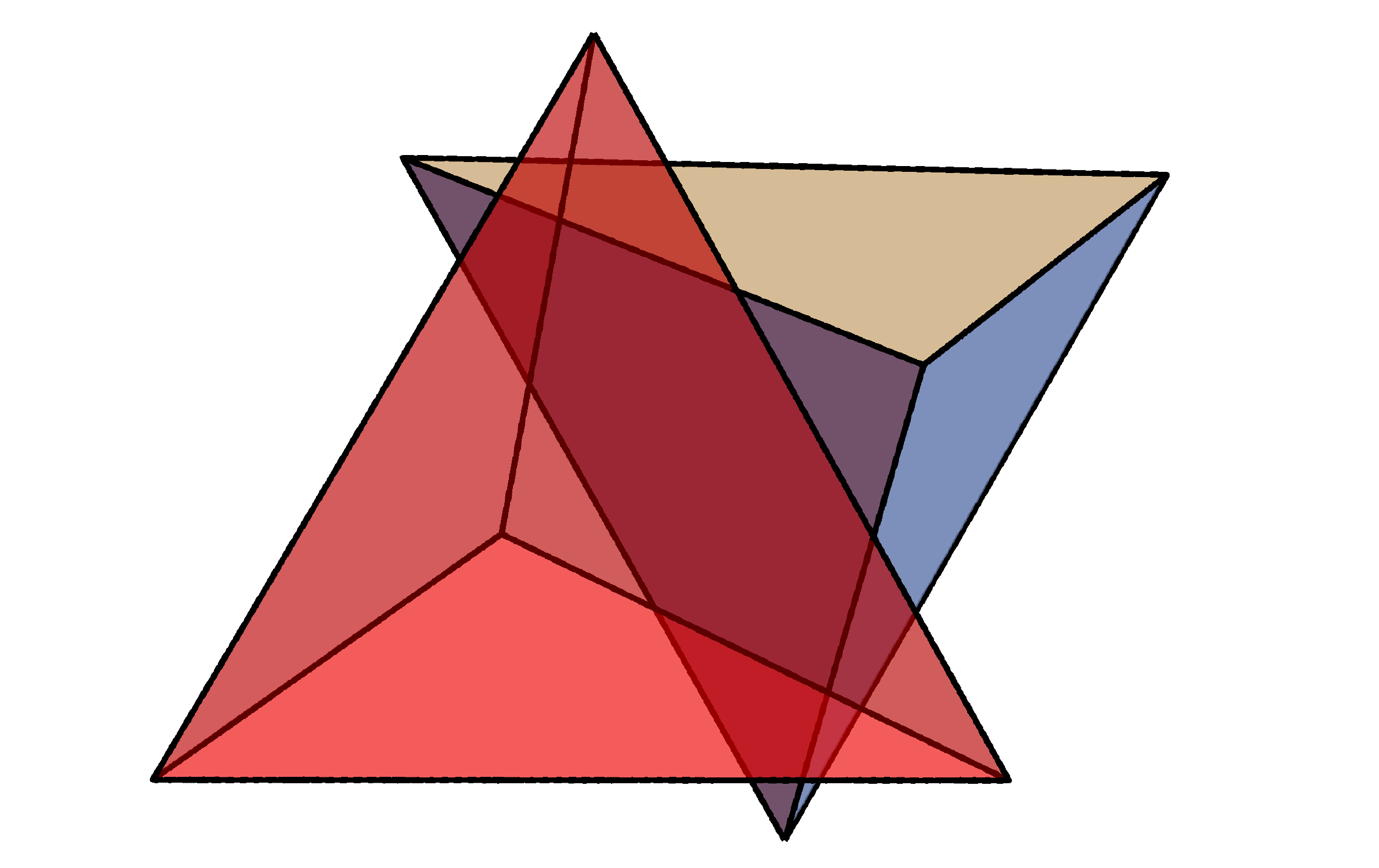}
\end{center}
\caption{A vector geometry associated to the dipole graph of Fig.~\ref{fig:dipole_graph}. This geometry consists of two regular tetrahedra. As a result we have shape-matching of adjacent faces and curvature along the edges with deficit angle $\delta=2\pi-2\arccos(1/3)$. In order to show that this Regge configuration is also a vector geometry it is sufficient to twist one of the two tetrahedra by the appropriate angle that sets all the normals of adjacent faces back-to-back.
}
\label{fig:dipole-vector-geometry}
\end{figure}

\subsection{Properties at fixed spins $(j_1,\ldots,j_{4})$}
We illustrate some properties of the Bell-network state on $\Gamma_{2}$ projected to fixed spins $j_{\ell}$. The state $\ket{\Gamma_2, \mathcal{B},j_\ell}$ is given by Eq.~(\ref{eq:gamma-fixed-j}).

The reduced density matrix for the subsystems $s$ is immediate to compute,
\begin{equation}
\rho_{s}=\text{Tr}_{i_{t}}\Big(\ket{\Gamma_2, \mathcal{B},j_\ell}\bra{\Gamma_2, \mathcal{B},j_\ell}\Big)=\frac{1}{\dim \mathcal{K}(j_{\ell})}\mathbbm{1}_{s}\,.
\label{eq:}
\end{equation}
As this density matrix is maximally mixed, the entanglement entropy of the subsystem $s$ is simply the $\log$ of the dimension of the intertwiner space at $s$,
\begin{equation}
S(\rho_{s})=-\text{Tr}(\rho_{s}\log \rho_{s})\,=\,\log \big(\dim \mathcal{K}(j_{\ell})\big)\,.
\label{eq:}
\end{equation}
The same happens for the subsystem $t$. On the other hand, for a dipole graph, the subsystem $st$ coincides with the full system which is in a pure state and therefore its entropy vanishes. As a result the mutual information of the subsystems $s$ and $t$ is
\begin{equation}
S(\rho_{s}\otimes \rho_{t}\|\rho_{st})=S(\rho_{s})+S(\rho_{t})-S(\rho_{st})\,=\,2 \log \big(\dim \mathcal{K}(j_{\ell})\big)\,.
\label{eq:}
\end{equation}
Now we focus on observables $\mathcal{O}_{s}$ and $\mathcal{O}_{t}$ which probe only the subsystem $s$ or $t$. We define expectation values and dispersions as usual,
\begin{equation}
\langle \mathcal{O}_{s} \rangle = \bra{\Gamma_2, \mathcal{B},j_\ell}  \mathcal{O}_{s} \ket{\Gamma_2, \mathcal{B},j_\ell}\,,\qquad \Delta \mathcal{O}_{s}=\sqrt{\langle \mathcal{O}_{s}{}^{2} \rangle-\langle \mathcal{O}_{s} \rangle^{2}}\,,
\label{eq:}
\end{equation}
and the norm of an observable as its largest eigenvalues, $\|\mathcal{O}\|=\max(o_{k})$. 
For any two observables in $s$ and $t$, the information theoretic inequality \cite{wolf2008area} 
\begin{equation}
\frac{1}{2}\left(\frac{\langle \mathcal{O}_{s}\,\mathcal{O}_{t}\rangle-\langle \mathcal{O}_{s}\rangle\,\langle \mathcal{O}_{t}\rangle}{\|\mathcal{O}_{s}\|\,\|\mathcal{O}_{t}\|}\right)^{2}\leq S(\rho_{s}\otimes \rho_{t}\|\rho_{st})\,.
\label{eq:}
\end{equation}
holds, with a non-vanishing right hand side for Bell-network states as checked above.

Let us now consider the observables $g_{ab}(s)=\vect{E}_{sa}\cdot\vect{E}_{sb}$ and $g_{ab}(t)=\vect{E}_{ta}\cdot\vect{E}_{tb}$ which measure the shape of the quantum tetrahedra $s$ and $t$. We define also the angle operator
\begin{equation}
\widehat{\cos\theta_{ab}}(s)\,=\,\frac{g_{ab}(s)}{\sqrt{g_{aa}(s)}\sqrt{g_{bb}(s)}}\,,
\label{eq:}
\end{equation}
which measures the cosine of the angle between the normals of the faces $a$ and $b$ of the quantum tetrahedron $s$. Its expectation value can be computed using standard techniques in recoupling theory and is given by
\begin{equation}
\big\langle \widehat{\cos\theta_{ab}}(s) \big\rangle = \frac{1}{\dim \mathcal{K}(j_{\ell})} \sum_{k=0}^{\dim \mathcal{K}(j_{\ell})-1} \frac{(k_{0}+k)(k_{0}+k+1)-j_a(j_a+1) - j_b(j_b+1)}{2 \sqrt{j_a(j_a+1)j_b(j_b+1)}} \, ,
\label{eq:average-angle}
\end{equation}
where $k_{0}=\max(|j_{1}-j_{2}|,|j_{3}-j_{4}|)$ for the observable $\big\langle \cos\theta_{12}(s) \big\rangle $ and defined by permutation for the other components. In the special case of spins all equal, $j_{\ell}=j_{0}$, we have $\dim \mathcal{K}(j_{0})=2j_{0}+1$, $k_{0}=0$ and average angle\footnote{
In the more general case of two pairs of equal spins, $j_1=j_2$ and $j_3=j_4$, the average values of the angle operators $\cos \theta_{ab}$ is:
\begin{equation}
\langle {{\cos \theta_{12}}} \rangle = -\frac{1}{3} \, ,\quad
\langle {{\cos \theta_{34}}}\rangle  = -1 +\frac{2}{3} \frac{j_1(j_1+1)}{j_3(j_3+1)} \, , \quad
\langle {{\cos \theta_{13}}}\rangle  = -\frac{1}{3} \sqrt{\frac{j_1(j_1+1)}{j_3(j_3+1)}} \, .
\label{eq:}
\end{equation}
}
\be
\big\langle \widehat{\cos\theta_{ab}}(s) \big\rangle =  - \frac{1}{3} \; ,\qquad\qquad  \qquad (\text{for}\;\;j_{1}=j_{2}=j_{3}=j_{4})
\label{eq:angle}
\ee
which coincides with the dihedral angle of a regular tetrahedron. For equal spins, the expectation value of the geometry of a quantum tetrahedron in a Bell state on $\Gamma_{2}$ is a regular tetrahedron with faces of area $A_{ab}=\mathfrak{a}_{0}\sqrt{j_{0}(j_{0}+1)}$. Therefore, in average the shapes of the two tetrahedra match: we don't just recover a vector geometry but in average a Regge geometry on $\Gamma_{2}$. Clearly, as implied by the Heisenberg uncertainty relations, the dispersion around the average cannot be vanishing. We find 
\begin{align}
\Delta(\cos\theta_{ab})=  \frac{1}{3 \sqrt{5}} \bigg(\sqrt{16 + \frac{3}{j_{0}} - \frac{3}{j_{0}+1}}\bigg) \xrightarrow{j_{0} \rightarrow \infty} \frac{4}{3 \sqrt{5}} \, ,
\end{align}
which remains finite in the limit of large spins.

Next we compute the correlation function for operators $g_{ab}(s)$ and $g_{ab}(t)$ describing observations of the same angle in the two tetrahedra:
\begin{equation}
\mathcal{C}_{abab}(s,t)=\frac{\langle g_{ab}(s)\,g_{ab}(t)\rangle-\langle  g_{ab}(s)\rangle\,\langle  g_{ab}(t)\rangle}{\| g_{ab}(s)\|\,\| g_{ab}(t)\|}= \frac{16}{45} \bigg(\frac{j(j+1) + 3/16}{j^{2}}\bigg) \xrightarrow{j \rightarrow \infty} \frac{16}{45} \, .
\label{eq:Cgg}
\end{equation}
The correlation function remains finite for $j \rightarrow \infty$, showing that the fluctuations of the geometry remain correlated in the semiclassical limit  of large spins.

\section{Bell-network state on the pentagram $\Gamma_{5}$}
\label{sec:Gamma5}
We describe the geometry of Bell-network states on the pentagram graph.
\subsection{The pentagram $\Gamma_{5}$}
\begin{figure}[t]
\begin{center}
 \includegraphics[scale=1]{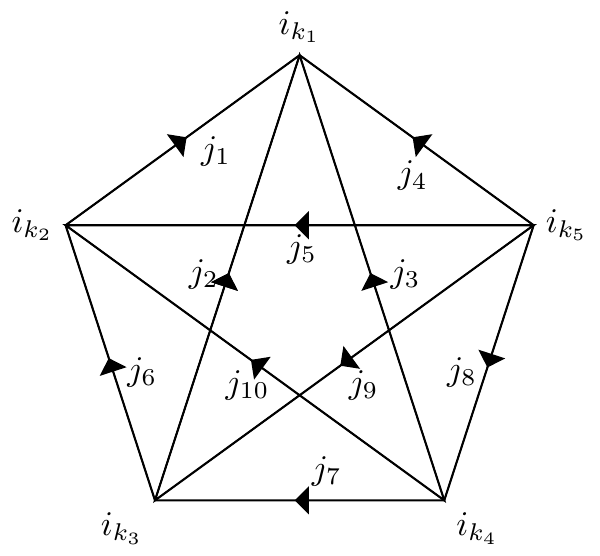}
\end{center}
\caption{Pentagram graph $\Gamma_5$. The graph is dual to a triangulation of the three-sphere with five tetrahedra. A spin-network state on $\Gamma_5$ is labeled by five intertwiners $i_{k_n}$ and ten spins $j_\ell$ attached to the nodes and links of the graph, respectively. The links are oriented according to $t(\ell)<s(\ell)$. \label{Fig:Pentagram}}
\end{figure}

The pentagram graph $\Gamma_5$ is formed by five nodes connected by ten links as shown in Fig.~\ref{Fig:Pentagram}. It is dual to a triangulation of $S^{3}$ with five tetrahedra. We label the nodes by an index $a=1,\dots,5$. An oriented link corresponds to an ordered pair $\ell=(a,b)$, where $a$ and $b$ are the source and target nodes, respectively. The space of states of loop quantum gravity on $\Gamma_5$ is the Hilbert space $\mathcal{H}_{\Gamma_5}=L^2[SU(2)^{10}/SU(2)^5]$. A spin-network basis state $\ket{\Gamma_5,j_{ab},i_a}$ is labeled by $10$ spins $j_{ab}$ and $5$ four-valent intertwiners $i_a$. It is a product state over intertwiners given by
\begin{equation}
\ket{\Gamma_5,j_{ab},i_a}=|i_{1}\rangle\,|i_{2}\rangle\,|i_{3}\rangle\,|i_{4}\rangle\,|i_{5}\rangle\,,
\label{eq:}
\end{equation}
where for instance the intertwiner $|i_{1}\rangle$ belongs to the space $\mathcal{K}_{1}=\text{Inv}(\mathcal{H}^{(j_{12})}\otimes\mathcal{H}^{(j_{13})}\otimes\mathcal{H}^{(j_{14})}\otimes\mathcal{H}^{(j_{15})})$. The Hilbert space $\mathcal{H}_{\Gamma_5}$ is a subspace of the bosonic Hilbert space used to construct squeezed vacua, $\mathcal{H}_{\Gamma_5} \subset \mathcal{H}_{bos}^{(40)}$, consisting of $40$ oscillators.

\subsection{Bell-network state on $\Gamma_{5}$}
A Bell-network state on $\Gamma_{5}$ is obtained projecting the associated squeezed vacuum, Eq.~\eqref{eq:Bell-net2}.  At fixed spins $j_{ab}$, we obtain the Bell-network state (explicitly derived in Appendix \ref{appendix:Hessian of the action for $15j$-symbol}):
\be
\label{eq:Bell_state_pentagram}
\ket{\Gamma_5,\mathcal{B},j_{ab}} = P_{\Gamma_5} \ket{\mathcal{B},j_{ab}} =  \sum_{i_a} \overline{\{15j\}(j_{ab},i_a)}  \left( \bigotimes_{k=1}^5 \ket{i_k} \right)  \, ,
\ee
where the $\{15j\}$-symbol is the contraction of the intertwiners $i_a$ along the graph $\Gamma_5$, the $SU(2)$ symbol of the graph $A_{\Gamma_{5}}(j_{ab},i_a)=\{15j\}(j_{ab},i_a)$. The expression of the state is then a superposition over spins
\begin{equation}
\ket{\Gamma_5,\mathcal{B},\lambda_{ab}}=\frac{1}{\mathcal{N}}\sum_{j_{ab}} \Big(\prod_{(ab)}  \sqrt{2j_{ab}+1}\, \lambda_{ab}^{2 j_{ab}}\Big)\;\ket{\Gamma_5,\mathcal{B},j_{ab}}\,,
\label{eq:}
\end{equation}
with parameters $|\lambda_{ab}|<1$ encoding the average area $A_{ab}$ and extrinsic angle $\Theta_{ab}$.

A useful representation of the state is obtained by  implementing the projection into the space of gauge-invariant states via coherent intertwiners, Eq.~(\ref{eq:coherent-intertwiner}). At fixed spins $j_{ab}$, the gauge invariant projector at nodes can be written as \cite{Livine:2007vk,Conrady:2009px}
\begin{equation}
P_{n}=\int d\mu(\vect{n}_{na})\,|\Phi_{n}(\vect{n}_{na})\rangle \langle \Phi_{n}(\vect{n}_{na})|\,,
\label{eq:}
\end{equation}
where $|\Phi_{n}(\vect{n}_{na})\rangle$ is the coherent intertwiner peaked on the polyhedron with normals $j_{na}\vect{n}_{na}$. The measure $d\mu(\vect{n}_{na})$ is gauge-invariant and can be expressed in terms of gauge-fixed normals via the shape parameters $(q_{ni},p_{ni})$ \cite{Conrady:2009px}. Using this expression, together with Eq.~(\ref{eq:coherent-intertwiner}), we find the expression of a Bell-network state as a superposition of coherent intertwiners: 
\begin{equation}
\ket{\Gamma_5,\mathcal{B},j_{ab}} =\int d\mu(\vect{n}_{na})\,\overline{\{15j\}(j_{ab},\vect{n}_{ab})}\,  \Big( \bigotimes_{n=1}^5 \ket{\Phi_{n}(\vect{n}_{na})} \Big)  \, .
\label{eq:Bell_state_pentagram_LS}
\end{equation}
In this formula, the quantity $\{15j\}(j_{ab},\vect{n}_{ab})$ is the familiar $\{15j\}$-symbol expressed in the coherent state basis, \cite{Barrett:2009as},
\begin{equation}
\{15j\}(j_{ab},\vect{n}_{ab})=\int_{SU(2)^5} \left( \prod_{n=1}^5 dg_n \right) \prod \limits_{a < b} \, \bra{j_{ab},\zeta \vect{n}_{ab}}U(g_a)^{\dagger}\,U(g_{b}) \ket{j_{ab},\vect{n}_{ba}}
\label{eq:15j-symbol} \, .
\end{equation}
Correlation functions on this state, $\langle g_{ab}(n)\,g_{cd}(n')\rangle-\langle  g_{ab}(n)\rangle\,\langle  g_{cd}(n')\rangle$, can be computed using the same techniques employed in the evaluation of the LQG propagator \cite{Bianchi:2006uf,Livine:2006it,Alesci:2008ff,Bianchi:2009ri,Bianchi:2011hp}.

\subsection{Large spin limit and vector geometries}
The large spin limit of the $\{15j\}$ symbol is well-studied \cite{Barrett:2009as,Dona:2017dvf}. The integral expression \eqref{eq:15j-symbol} can be analyzed via saddle point methods in the limit of large spins by rescaling $j_{ab} \rightarrow \lambda j_{ab}$, letting $\lambda \rightarrow \infty$ and studying its asymptotic expansion in $1/\lambda$. In fact \eqref{eq:15j-symbol} can be written as
\begin{equation}
\{15j\}(\lambda j_{ab},\vect{n}_{ab})=\int\Big(\prod_{n=1}^5 dg_n \Big) e^{\lambda S(j_{ab},\vect{n}_{ab})}
\label{eq:15j-action} \, ,
\end{equation}
with the complex function $S(j_{ab},\vect{n}_{ab})$ given by
\begin{equation}
\textstyle S(j_{ab},\vect{n}_{ab})=\sum_{ab}2j_{ab}\log\,\bra{\frac{1}{2},\zeta \vect{n}_{ab}}g_a^{-1}g_{b} \ket{\frac{1}{2},\vect{n}_{ba}}\,.
\label{eq:action}
\end{equation}
The integral is dominated by saddle points where the real part of this function vanishes,
\begin{equation}
0=\text{Re}\,S(j_{ab},\vect{n}_{ab})=\sum_{ab}2j_{ab}\log \frac{1-(R_a\vect{n}_{ab})\cdot(R_{b}\vect{n}_{ba})}{2}\,.
\label{eq:}
\end{equation}
Here $R_{a}=D^{(1)}(g_{a})$ is the adjoint representation of the $SU(2)$ group element $g_{a}$. Clearly, this is equivalent to the requirement for the existence of a vector geometry, Eq.~(\ref{eq:normal-matching}),
\begin{equation}
R_a \vect{n}_{ab} = - R_b \vect{n}_{ba} \,.
\label{eq:}
\end{equation}
If the normals $\vect{n}_{ab}$ do not describe a vector geometry, the symbol $\{15j\}(\lambda j_{ab},\vect{n}_{ab})$ is exponentially suppressed in the limit of large spins,
\begin{equation}
\{15j\}(\lambda j_{ab},\vect{n}_{ab})\Big|_{\text{non-vec}}=o(\lambda^{-n})\qquad\forall n>0\,.
\label{eq:}
\end{equation}
On the other hand, if the set of normals $\vect{n}_{ab}$ describes a vector geometry, the asymptotic behavior of the $\{15j\}$ symbol is not exponentially suppressed and is given by the expression
\begin{equation}
\{15j\}(\lambda j_{ab},\vect{n}_{ab})\Big|_{\text{vec}} = \bigg(\frac{2 \pi}{\lambda}\bigg)^6 \frac{2^4}{(4 \pi)^8} \sum_{\sigma}\frac{e^{i \lambda S_\sigma}}{\sqrt{\mathrm{det} \, H_\sigma}}\;\;+\textstyle O(\lambda^{-7})\,.
\label{eq:15jsaddle}
\end{equation}
where $\sigma$ stands for the set of saddle points $g_{a}^{(\sigma)}$ that dominate the integral (\ref{eq:15j-action}), $S_{\sigma}$ is the imaginary part of the function (\ref{eq:action}) evaluated at these saddle points and $H_\sigma$ its Hessian.

Using these well-known results on the asymptotics of the $\{15j\}$ symbol together with Eq.~(\ref{eq:Bell_state_pentagram_LS}), we conclude that in the large spin limit a Bell-network state represents a uniform superposition over vector geometries:
\begin{equation}
\ket{\Gamma_5,\mathcal{B},j_{ab}} \approx\bigg(\frac{2 \pi}{\lambda}\bigg)^6 \frac{2^4}{(4 \pi)^8} \sum_{\sigma}\int_{\text{vec-geom}}\hspace{-2em} d\mu(\vect{n}_{ab})\, \frac{e^{-i \lambda S_\sigma}}{\sqrt{\mathrm{det} \, H_\sigma}}\,  \Big( \bigotimes_{n=1}^5 \ket{\Phi_{n}(\vect{n}_{na})} \Big)  \, .
\label{eq:vec-sup}
\end{equation}
Therefore, at fixed large spins on $\Gamma_{5}$, a Bell-network state represents five entangled polyhedra with glued adjacent faces.

\subsection{Gluing tetrahedra in the 4-1 configuration}
The relative weight of different vector geometries in the expression Eq.~(\ref{eq:vec-sup}) is determined by the Hessian $H_{\sigma}$. The saddle points $\sigma$ appearing Eq.~(\ref{eq:15jsaddle}) are vector geometries classified as follows \cite{Barrett:2009as,Dona:2017dvf}: 
\begin{itemize}
\item[-] If at fixed $\vect{n}_{ab}$ there are two inequivalent solutions to the saddle point equations, then the data $\{j_{ab},\vect{n}_{ab}\}$ are necessarily shape-matched, (SM-2);
\item[-] When there is only one set of solutions to the critical equations up to equivalence, the boundary data can be shape-matched (SM-1) or normal-matched (NM-1).
\end{itemize}
As Regge geometries are a subset of the space of vector geometries, it is interesting to study what is their relative weight.  To this end, we choose a specific configuration of spins and explore the dependence of $|\mathrm{det} H|^{1/2}$ on the normals $\vect{n_{ab}}$.

\begin{figure}[htbp]
\begin{center}
 \includegraphics[scale=1.25]{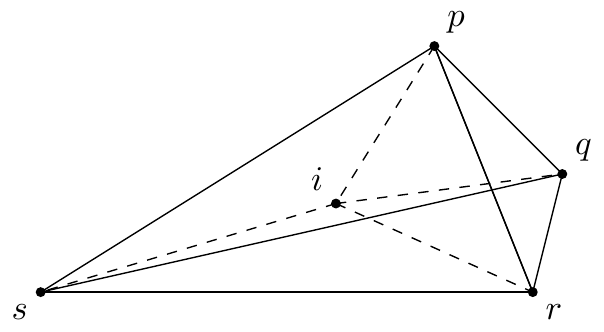}
\end{center} 
\caption{Pachner move 1-4. A single tetrahedron is divided into a gluing of four tetrahedra by the introduction of a new vertex $i$ in its interior.}
\label{fig:1-4}
\end{figure}

\begin{figure}[t]
\begin{center}
 \includegraphics[scale=0.5]{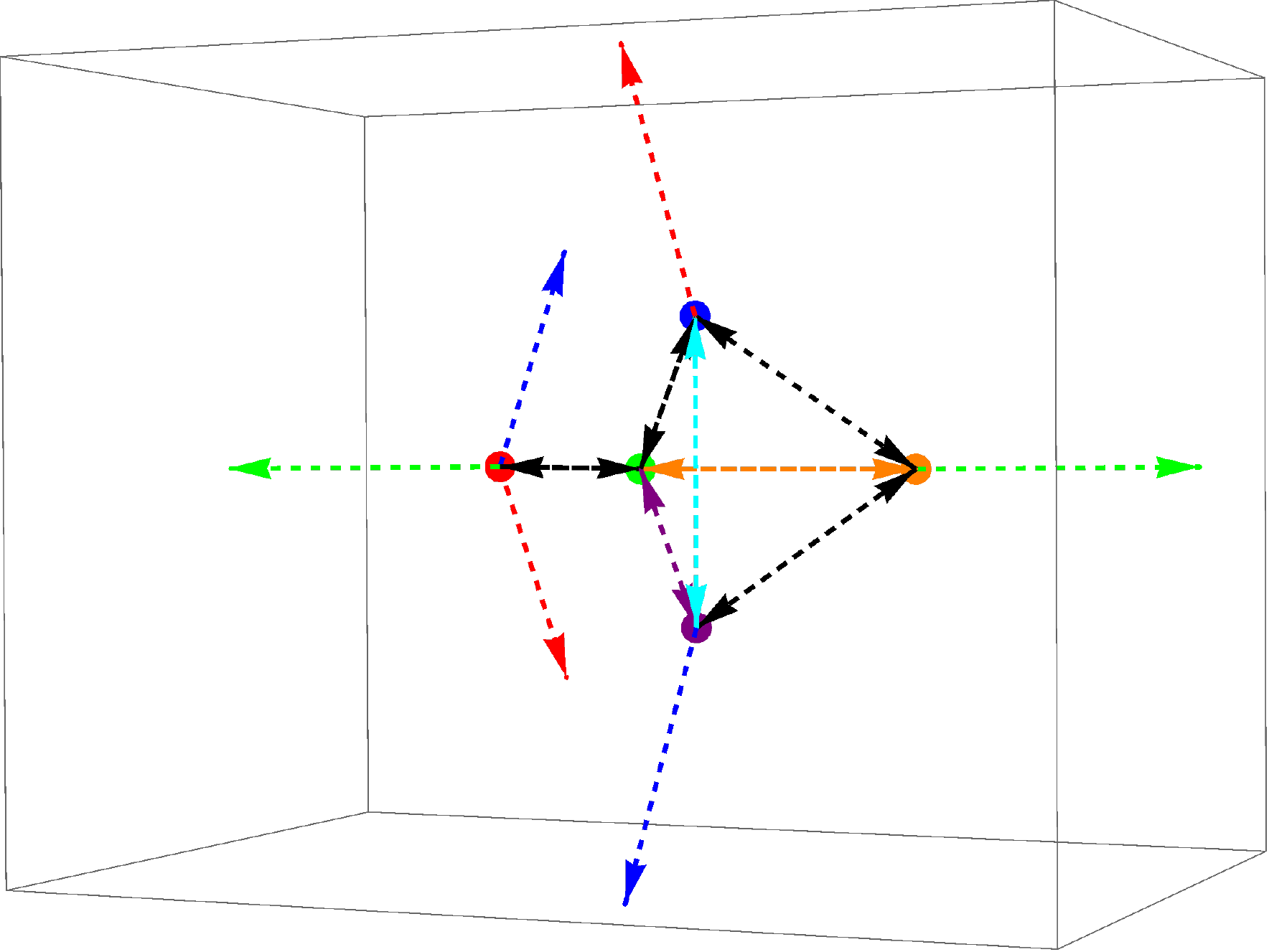}
\end{center} 
\caption{Graphical representation of the 1-4 vector geometry shown in Fig.~\ref{fig:vector-geometry}. The representation is generated by using maximal tree method. Each colored point labels a node of the pentagram. The set of normals at each node satisfies the closure condition \eqref{eq:closure}. Dashed black arrows stand for the anti-aligned normals which connect the neighboring nodes: $\vect{n}_{a (a+1)}=-\vect{n}_{(a+1) a}$, $a=1,\cdots,4$. Dashed colored arrows stand for the remaining normals with $\vect{n}_{ab}=-\vect{n}_{ba}$ and $\big\{(a,b)=1,\cdots,5 \bigm| a < b \;\;\text{and}\;\, a \neq b-1\big\}$.}
\label{fig:vector_geometry_graphical_v2}
\end{figure}

By choosing an arbitrary node $a$ of the pentagram, the triangulation associated to $\Gamma_5$ can be seen as the gluing of two pieces related by a Pachner move 1-4: a single tetrahedron $a$ and the polyhedron with four tetrahedra obtained from it by the inclusion of an internal point (see Fig.~\ref{fig:1-4}). Now, if the data $\{j_{ab},\vect n_{ab}\}$ is that of a shape-matched $3d$ Euclidean geometry, then these two pieces can be isometrically embedded in $\mathbb{R}^3$. Reversing the procedure, we obtain a method for constructing shape-matched configurations: we apply a Pachner move to a tetrahedron embedded in $\mathbb{R}^3$, and then just read off the boundary data from the explicit embedding in order to build an Euclidean $3d$ geometry for the triangulation of $S^{3}$. This allows us, in particular, to determine the coordinates of the shape-matched configurations for a given parametrization of the solutions of the saddle point conditions. The whole procedure can equally well be based on the Pachner move 2-3. In Appendix~\ref{appendix:Shape-matching configurations for a 4-simplex}, we construct explicit shape-matched configurations (SM-1) for 1-4 and 2-3 vector geometries (Fig.~\ref{fig:vector_geometry_graphical_v2}) and compute the amplitude of the corresponding $\{15j\}$-symbols by computing $|\mathrm{det} H|^{1/2}$. 

In \cite{Dona:2017dvf} it is shown that, on $\Gamma_{5}$, vector geometries with fixed spins can be parametrized in terms of five independent variables $\phi$. The five tetrahedra are represented in the Kapovich-Millson phase space as 4-sided polygons in $\mathbb{R}^3$ with edge vectors $j_{ab} \, \vect{m}_{ab}$, and the geometry is completely determined for a given set of spins if the face normals $\vect{m}_{ab}$ are written in terms of the five independent shape variables $\phi$. Four of these shape variables are gauge-invariant $3d$ dihedral angles computed from the squared length of diagonals of the polygons:
\begin{align}
\label{normals_polygon2}
(\vect{m}_{ca} + \vect{m}_{cd}) \cdot(\vect{m}_{ca} +  \vect{m}_{cd})  &= 2 \, (1  + \cos \phi_{ca,cd}) \\
(-\vect{m}_{ab} + \vect{m}_{ca}) \cdot(-\vect{m}_{ab} + \vect{m}_{ca})  &= 2 \, (1 + \cos \phi_{ab,ca}) \\
(-\vect{m}_{db} +  \vect{m}_{cd})\cdot (- \vect{m}_{db} +  \vect{m}_{cd}) &=2 \, (1 + \cos \phi_{db,cd}) \\ 
(\vect{m}_{ab} +  \vect{m}_{db})\cdot ( \vect{m}_{ab} + \vect{m}_{db})  &= 2 \, (1 +  \cos \phi_{ab,db}) \, ,
\end{align}
and the remaining variable is a gauge-dependent quantity, an angle between faces at distinct tetrahedra,
\bea
\label{normals_polygon1}
 (\vect{m}_{ab} + \vect{m}_{cd})  \cdot (\vect{m}_{ab} + \vect{m}_{cd})  =  2 \, (1+ \cos \phi_{cd,ab}) \, ,
 \eea
computed from the squared length of the diagonal of the parallelogram ($\vect{m}_{cd}, \vect{m}_{dc}, \vect{m}_{ab}, \vect{m}_{ba}$).

We are interested in the 1-4 geometry obtained by the application of the Pachner move 1-4 to a regular tetrahedron. Accordingly, the exterior normals ($\vect{m}_{af},\vect{m}_{bf}, \vect{m}_{cf},\vect{m}_{df}$) are fixed to match the normals ($\vect{n}_{af},\vect{n}_{bf}, \vect{n}_{cf},\vect{n}_{df}$) of an explicit embedding of the regular tetrahedron in $\mathbb{R}^3$ (as described in Appendix \ref{appendix:Shape-matching configurations for a 4-simplex}). In addition, the spin $j_{in}$ of the interior faces is related to the spin $j_{out}$ of the exterior faces by $ j_{\mathrm{in}}=j_{\mathrm{out}} /\sqrt{6}$. This identity cannot hold for semi-integer spins, but can be arbitrarily well-approximated for large spins, which is the regime we are interested in. The two independent closure conditions for the interior tetrahedra read:
\bea
\vect{m}_{cb} &= - \frac{1}{j_{\mathrm{in}}} (j_{\mathrm{out}} \, \vect{n}_{fb} + j_{\mathrm{in}} \, \vect{m}_{db} + j_{\mathrm{in}} \, \vect{m}_{ab} ) \, , \\
\vect{m}_{ad} &= - \frac{1}{j_{\mathrm{in}}} (j_{\mathrm{out}} \, \vect{n}_{fd} + j_{\mathrm{in}} \, \vect{m}_{bd} + j_{\mathrm{in}} \, \vect{m}_{cd} ) \, .
\eea
Overall we obtain a 1-4 vector geometry with fixed spins from five shape variables where the non-gauge invariant quantity $\phi_{cd,ab}$ is the angle between normals of different tetrahedra and the other four variables are the dihedral angles $\phi_{ca,cd},\phi_{ab,ca},\phi_{db,ab},\phi_{ab,db}$.\footnote{The 2-3 vector geometries are too constrained for our purposes. We could start with two regular tetrahedra glued back to back and let the interior geometry of the piece formed by three tetrahedra free. A simple counting argument shows, however, that this interior geometry is rigidly fixed. In general, a vector geometry is described by 40 parameters in 20 normalized vectors. We have 10 equations from the solutions to the critical equations. Two tetrahedra are constructed to be regular such that 7 normals are fixed in the vectorial geometry. There are also 3 closure conditions for the three free tetrahedra. The total number of constraints is $10 \times 2 + 7 \times 2 + 3 \times 2 = 40$ which is exactly the total number of parameters in a vector geometry. Therefore, the vectorial equations are too constrained to allow for configurations other than the shape-matched one.}

We now proceed to compare the contributions of shape-matched and normal-matched configurations to the $\{15j\}$-symbols for the chosen spin configuration. The shape-matched configuration belongs to the SM-1, therefore it is enough to consider SM-1 and NM-1 contributions. In the limit of large spins, the $\{15j\}$-symbols have the asymptotic form given in Eq.~\eqref{eq:15jsaddle}. Therefore, in the asymptotic limit the ratio between the $\{15j\}$-symbols for normal-matched and shape-matched configurations is given by $\sqrt{|\det H_0|/|\det H|}$, where $H_0$ is the Hessian of the shape-matched configuration (whose coordinates are derived in Appendix \ref{appendix:Shape-matching configurations for a 4-simplex}).

Using the parametrization introduced above, we first sample points of a subset of 1-4 vector geometries including the shape-matched solution.  For this purpose, we first express $(\vect{m}_{cd},\vect{m}_{ab})$ in terms of four parameters as
\bea
\vect{m}_{cd}^T &= (\sin \theta_{cd} \cos \phi_{cd}, \sin \theta_{cd} \sin \phi_{cd}, \cos \theta_{cd}) \, , \\
\vect{m}_{ab}^T &= (\sin \theta_{ab} \cos \phi_{ab}, \sin \theta_{ab} \sin \phi_{ab}, \cos \theta_{ab}) \, .
\eea
The other two interior normals in \eqref{normals_polygon2} are fixed to the shape-matched solutions $\vect{m}_{ca} = \vect{n}_{ca}$, $\vect{m}_{ca} = \vect{n}_{db}$, and $\vect{m}_{cb},\vect{m}_{ad}$ are determined by the following closure conditions:
\bea
\label{eq:closure1}
\vect{m}_{cb} &= - \frac{1}{j_{\mathrm{in}}} (j_{\mathrm{out}}  \vect{n}_{fb} + j_{\mathrm{in}} \vect{n}_{db} + j_{\mathrm{in}}  \vect{m}_{ab}) \, , \\
\vect{m}_{ad} &= - \frac{1}{j_{\mathrm{in}}} (j_{\mathrm{out}}  \vect{n}_{fd} + j_{\mathrm{in}} \vect{n}_{bd} + j_{\mathrm{in}}  \vect{m}_{cd} ) \, .
\eea
Then we look for solutions within the interval $[0, 2 \pi]$ of $\theta \equiv \theta_{cd}=\theta_{ab}$. This choice of parametrization leads to equalities among the gauge-invariant quantities:
\bea
\cos \phi_{ab,ca} - \cos \phi_{ab,db} = \cos \phi_{ca,cd} - \cos \phi_{db,cd} = \sqrt{2} \cos \theta \, ,
 \eea
where all these gauge-invariant quantities are exactly equal to $1/2$ at the shape-matched configuration. The corresponding solutions for $\phi_{ab}, \phi_{cd}$ are obtained by the normalization conditions on $\vect{m}_{cb},\vect{m}_{ad}$.

\begin{figure}[t]
\begin{center}
 \includegraphics[scale=0.45]{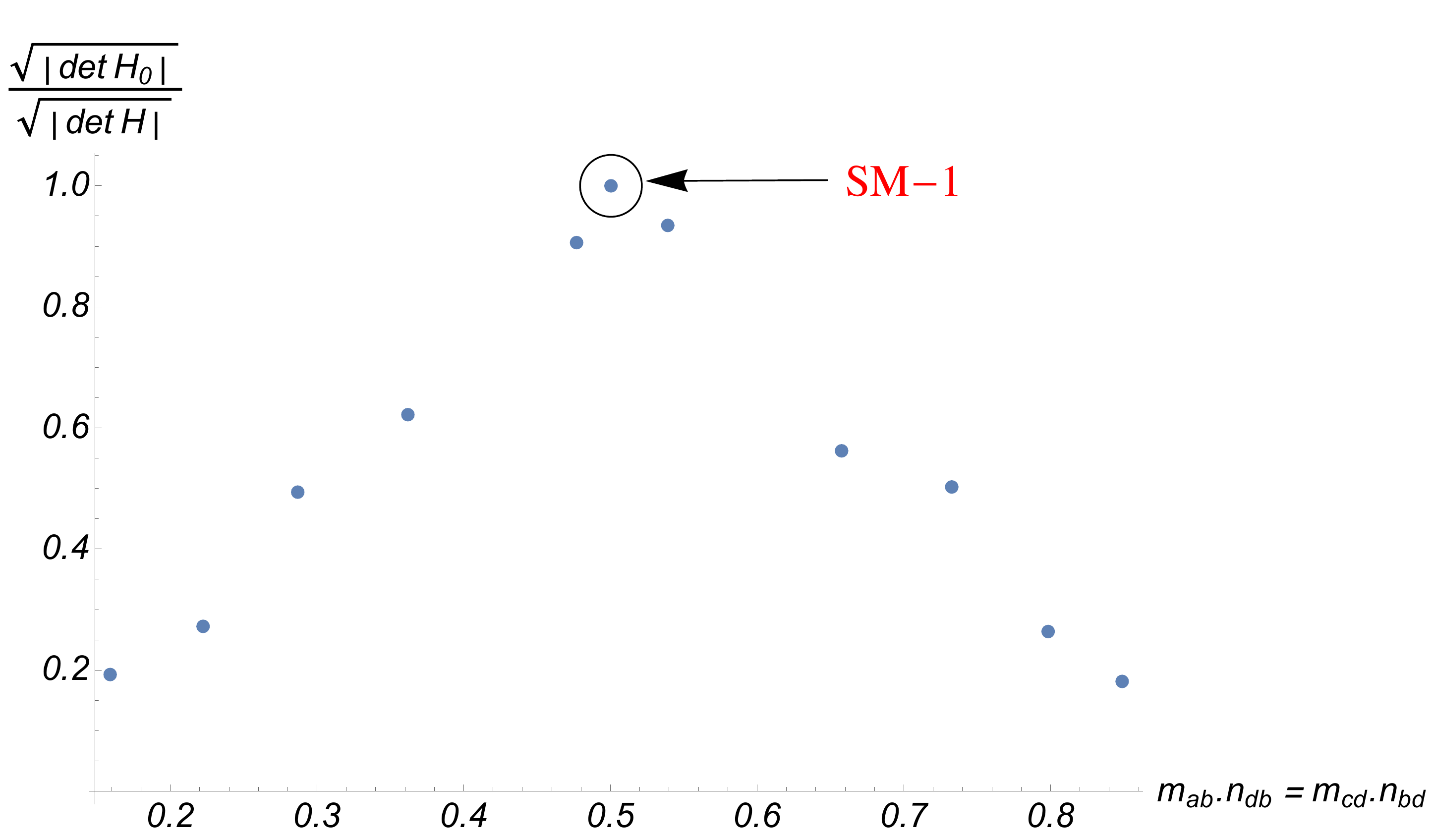}
\end{center} 
\caption{Sampling of exact relative magnitudes $(|\mathrm{det} H_0|^{1/2}_{1-4})/(|\mathrm{det} H|^{1/2}_{1-4})$ in terms of the cosine of the gauge-invariant 3d dihedral angles $ \phi_{ab,db}$ and $ \phi_{cd,bd}$. The relative magnitude reaches its maximum at the shape-matched configuration (SM-1) with $\vect{n}_{ab} \cdot \vect{n}_{db} = \vect{n}_{cd} \cdot \vect{n}_{bd} = 1/2$.}
\label{fig:dethes_plot}
\end{figure}

Now we can plot the relative magnitude of $|\mathrm{det} H|^{-1/2}$ with respect to the shape-matched configuration for a sample of exact vector geometries. The results are displayed in Fig.~\ref{fig:dethes_plot}. We find that the shape-matched configuration SM-1 is present in the superposition (\ref{eq:vec-sup}) and interestingly it gives the contribution with the largest amplitude. Therefore the Bell-network state on $\Gamma_{5}$ describes a superposition of glued tetrahedra with a significant contribution given by a Regge geometry.

\section{Summary and discussion}
\label{sec:summary}
In loop quantum gravity, spin-network basis states are eigenstates of local operators. These operators measure the quantum geometry of nodes and links of a spin-network graph. As a result, nearby nodes in a spin-network basis state are un-entangled: their geometry has uncorrelated fluctuations. At the classical level this behavior corresponds to a twisted geometry --- the geometry of a collection of polyhedra with uncorrelated shapes. In this paper we introduced a class of states with nearest-neighbors correlations that guarantee that neighboring polyhedra are glued at adjacent faces. We dub these states \emph{Bell-network states} as they are built by entangling nearby nodes in a way that generalizes the spin-spin correlations in a Bell state of two spin-$1/2$ particles.

Bell-network states are built using squeezed vacuum techniques and are labeled by $L$ complex numbers $\lambda_{\ell}$, one per link of a graph $\Gamma$. The modulus and phase of the parameter $\lambda_{\ell}$ encode the average area and extrinsic angle $(A_{\ell},\Theta_{\ell})$ of a link of the graph. The quantum geometry of nearby nodes is entangled in such a way that the normals to adjacent faces of neighboring polyhedra are always back-to-back, i.e., $\vect{n}_{s(\ell)}=-\vect{n}_{t(\ell)}$. This condition guarantees that the planes of the adjacent faces can be glued to each other. We note that this condition does not impose that the geometry is flat or that the shape of the faces matches. The geometric structure that arises has been previously studies in the spinfoam literature and is called a \emph{vector geometry}. Remarkably, a Bell-network state is not peaked on any single vector geometry. The picture that arises from our analysis is that, at fixed spins, a Bell-network state can be understood as a uniform superposition over all vector geometries. This behavior reflects the one of a Bell state of two spin-$1/2$ particles,
\begin{equation}
|\mathcal{B}\rangle =\frac{|\!\uparrow\rangle_{s}\,|\!\downarrow\,\rangle_{t}\,-\,|\!\downarrow\rangle_{s}\,|\!\uparrow\,\rangle_{t}}{\sqrt{2}}=\sqrt{2}\int \frac{d^{2}\vect{n}}{4\pi}\;|\vect{n}\rangle_{s}\,|\text{-}\vect{n}\rangle_{t}\,,
\label{eq:Bell}
\end{equation}
which can also be understood as a uniform superposition of back-to-back spins $|\vect{n}\rangle_{s}\,|\text{-}\vect{n}\rangle_{t}$ over all directions $\vect{n}$.

For a generic graph $\Gamma$, a Bell-network state is given by the expression
\begin{equation}
\textstyle |\Gamma,\mathcal{B},\lambda_{\ell}\rangle =P_{n}\,\exp\Big(\sum_{\ell} \lambda_{\ell}\, \epsilon_{AB}\,a^{\dagger A}_{s(\ell)}a^{\dagger B}_{t(\ell)}\Big)\;|0\rangle_{\Gamma}\label{eq:Bell-net-summary}\,,
\end{equation}
where $P_{n}$ is the gauge-invariant projector at nodes of the graph and the exponential of $a^{\dagger A}_{s(\ell)}a^{\dagger B}_{t(\ell)}$ squeezes links of the graph creating entangled pairs at its endpoints. The state can be expanded on a overcomplete basis of states consisting of uncorrelated intertwiners $|\Phi_{n}(\{\vect{n}\})\rangle $ peaked on a polyhedron with fixed face areas and normals:
\begin{equation}
|\Gamma,\mathcal{B},\lambda_{\ell}\rangle =\sum_{j_{\ell}}\sqrt{2j_{\ell}+1}\, \lambda_{\ell}^{2 j_{\ell}}\int\! d\mu(\{\vect{n}\}) \;A_{\Gamma}(j_{\ell},\{\vect{n}\})\;\bigotimes_{n}|\Phi_{n}(\{\vect{n}\})\rangle 
\label{eq:}
\end{equation}
where the amplitude $A_{\Gamma}(j_{\ell},\{\vect{n}\})$ is the $SU(2)$ symbol associated to the graph $\Gamma$ and expressed in a coherent basis. In the large spin limit it is know that this amplitude suppresses exponentially all configurations of normals except the ones for which there exists a choice of rotation matrices $R_{n}$ for which \cite{Dona:2017dvf}
\begin{equation}
R_{s(\ell)}\,\vect{n}_{s(\ell)}=-R_{t(\ell)}\,\vect{n}_{t(\ell)}\,.
\label{eq:}
\end{equation}
This is exactly the defining condition of a classical vector geometry.

We studied in detail properties of Bell-network states on simple graphs. On the dipole graph consisting of two tetrahedra we analyzed the correlation functions and showed that the shape of the two quantum tetrahedra are correlated in such a way that the two are always glued. In the case of the pentagram graph consisting of five tetrahedra we used the relation to $SU(2)$ $\{15j\}$-symbols, where vector geometries are known to appear in the large spin limit as saddle point configurations, to show explicitly the gluing of tetrahedra in the $4-1$ configuration.

The results presented show clearly the role of entanglement in the gluing of quantum regions of space. Bell-network states encode nearest-neighbor correlations in quantum polyhedra that enforce the gluing conditions. Long-range correlations are unconstrained and can be included via quantum squeezing as discussed in \cite{Bianchi:2016tmw,Bianchi:2016hmk}. As shown in this paper, the mutual information of the quantum geometry of nearby nodes provides a powerful tool to quantify the strength of correlations. With the inclusion of long-range correlations, these same techniques can be extended to the study of entanglement between large regions of space consisting of many degrees of freedom, a calculation of relevance for the identification of the semiclassical regime of loop quantum gravity.

\section*{Acknowledgements}
We thank Pietro Don\`a, Hal Haggard and Wolfgang Wieland for useful discussions on vector geometries. The work of EB is supported by the NSF grant PHY-1404204. The work of BB is supported by the NSF grant PHY-1417385. NY acknowledges support from CNPq, Brazil and the NSF grant PHY-1505411.

\appendix

\section{Mutual information and correlations in the Bell state of spin $j$}
\label{sec:mutual-info}
We compute the mutual information and the spin correlations in the Bell state of spin $j$,
\begin{equation}
 |\mathcal{B},j\rangle  =\frac{1}{\sqrt{2j+1}}\sum_{m=-j}^{+j} (-1)^{j-m}\; \ket{j, m}_{s} \ket{j, -m}_{t}\,.
\label{eq:}
\end{equation}
As the state is pure and the reduced density matrix $\rho_{s}=\frac{1}{2j+1}\mathds{1}$ is maximally mixed, we have that the mutual information in the source and target of the Bell state is
\begin{equation}
S(\rho_{s}\otimes \rho_{t}\|\rho_{st})=S(\rho_{s})+S(\rho_{t})-S(\rho_{st})\,=\,2\log(2j+1)\,.
\label{eq:}
\end{equation} 
The spin expectation values on the Bell state vanish
\begin{equation}
\langle \mathcal{B},j| J_{s}^{i}|\mathcal{B},j\rangle=0\,,\qquad \langle \mathcal{B},j| J_{t}^{i}|\mathcal{B},j\rangle=0\,,
\label{eq:}
\end{equation}
and the spin-spin correlation functions are
\begin{equation}
C^{ij}=\langle \mathcal{B},j| J_{s}^{i}\,J_{t}^{j}|\mathcal{B},j\rangle=-\frac{j(j+1)}{3}\delta^{ij}\,.
\label{eq:}
\end{equation}
Note that the spin operator is a bounded operator with norm $\|J^{i}\|=j$. Therefore the information-theoretic inequality
\begin{equation}
\frac{(C^{ij})^{2}}{2 \|J^{i}_{s}\|^{2}\,\|J^{i}_{t}\|^{2}}\;\leq\; S(\rho_{s})+S(\rho_{t})-S(\rho_{st})\,
\label{eq:}
\end{equation}
applies, as can also be checked explicitly noticing that $\frac{(j+1)^2}{18\, j^2}\leq 2 \log (2j+1)$.

\section{Derivation of the Bell states on a pentagram graph}
\label{appendix:Hessian of the action for $15j$-symbol}

The projection of $\ket{\mathcal{B}_{\gamma}} \in \mathcal{H}_{bos}$ to the space of spin network states $\mathcal{H}_{\Gamma_5} \subset \mathcal{H}_{bos}$ is
\be
\ket{\Gamma_5,\mathcal{B}_{\gamma}} = P_{\Gamma_5} \ket{\mathcal{B}_{\gamma}} \, .
\ee
Fixing the spins, we obtain a Bell state with determined spins:
\be
\ket{\Gamma_5,\mathcal{B}_{\gamma},j_{ab}} = P_{\Gamma_5} \ket{\mathcal{B}_{\gamma},j_{ab}} \, ,
\ee
where 
\be
\ket{\mathcal{B}_{\gamma},j_{ab}} = \bigotimes_{a<b} \sum_{m_{ab}} (-1)^{j_{ab}-m_{ab}} \ket{j_{ab},m_{ab}} \otimes \ket{j_{ba},-m_{ab}} \, .
\label{eq:bell-j}
\ee
We adopt the convention that a ket $\ket{j_{ab},m_{ab}}$ with spin $j_{ab}$ lives at the endpoint $a$ of the link $\overline{ab}$. The state $\ket{\mathcal{B}_{\gamma},j_{ab}}$ is by construction area-matched. The projection to the space of gauge-invariant states is easily constructed using the orthonormal intertwiner basis labeled by the virtual spins $i_a$:
\be
P_{\Gamma_5} =  \bigotimes_a P_a \, , \qquad P_a = \sum_{i_a} \ket{i_a} \bra{i_a} \, ,
\ee
where $\ket{i_a} \in \mathcal{H}_a=\text{Inv}[\bigotimes_{b:a\neq b} V_{j_{ab}}]$. 

We represent the orthonormal intertwiners in the magnetic basis as:
\be
\ket{i_1} = \sum_m [i_1]^{m_{12} m_{13} m_{14} m_{15}} \ket{j_{12},m_{12}} \otimes \ket{j_{13},m_{13}} \otimes \ket{j_{14},m_{14}} \otimes \ket{j_{15},m_{15}}\, ,  
\ee
and similarly for the other values of $a$. The dual bases with respect to the standard Hilbert space inner product and to the $\epsilon$ bilinear form are:
\begin{align*}
\bra{i_1} &= \sum_m \bar{[i_1]}^{m_{12} m_{13} m_{14} m_{15}} \bra{j_{12},m_{12}} \otimes \bra{j_{13},m_{13}} \otimes \bra{j_{14},m_{14}} \otimes \bra{j_{15},m_{15}}\, ,  \\
[i_1\vert &= \sum_m \left[ \prod_b (-1)^{j_{1b}-m_{1b}} \right][i_1]^{-m_{12},- m_{13},- m_{14},-m_{15}} \bra{j_{12},m_{12}} \otimes \bra{j_{13},m_{13}} \otimes \bra{j_{14},m_{14}} \otimes \bra{j_{15},m_{15}}\, .
\end{align*}
The antilinear time-reversal operator is defined as $\zeta \ket{i_1} = \vert i_1]$, with
\[
\vert i_1] = \sum_m \left[ \prod_{b\neq 1} (-1)^{j_{1b}-m_{1b}} \right]\bar{[i_1]}^{-m_{12},- m_{13},- m_{14},-m_{15}} \ket{j_{12},m_{12}} \otimes \ket{j_{13},m_{13}} \otimes \ket{j_{14},m_{14}} \otimes \ket{j_{15},m_{15}}\, .
\]

The projector $P_{\Gamma_5}$ can be implemented node by node. We first represent the state $\ket{\mathcal{B}_{\gamma},j_{ab}}$ (before the projection) as a superposition of tensor products of node states in the form:
\be
\ket{\mathcal{B}_{\gamma},j_{ab}} = \sum_{m_{ab}} \bigotimes_a \ket{\mathcal{B}_{\gamma_a}(m_{ab})} \, .
\label{eq:node-factorization}
\ee
Then we project each of the components in the above expansion to the gauge-invariant subspace. The representation \eqref{eq:node-factorization} is not unique, and we can choose, for instance:
\begin{align*}
\ket{\mathcal{B}_{\gamma_1}} &= \ket{j_{12},m_{12}} \otimes \cdots \otimes \ket{j_{15},m_{15}} \, , \\
\ket{\mathcal{B}_{\gamma_2}} &= (-1)^{j_{12}-m_{12}} \ket{j_{21},-m_{12}} \otimes \ket{j_{23},m_{23}} \otimes \ket{j_{14},m_{14}} \otimes \ket{j_{15},m_{15}} \, , \\
\ket{\mathcal{B}_{\gamma_3}} &= \left[ \prod_{b<3} (-1)^{j_{b3}-m_{b3}} \right] \ket{j_{31},-m_{13}} \otimes \ket{j_{32},-m_{23}} \otimes \ket{j_{34},m_{34}} \otimes \ket{j_{35},m_{35}} \, , \\
\ket{\mathcal{B}_{\gamma_4}} &= \left[ \prod_{b<4} (-1)^{j_{b4}-m_{b4}} \right]  \ket{j_{41},-m_{14}} \otimes \ket{j_{42},-m_{24}} \otimes \ket{j_{43},-m_{34}} \otimes \ket{j_{45},m_{45}} \, , \\
\ket{\mathcal{B}_{\gamma_5}} &= \left[ \prod_{b<5} (-1)^{j_{b5}-m_{b5}} \right] \ket{j_{51},-m_{15}} \otimes \ket{j_{52},-m_{25}} \otimes \ket{j_{53},-m_{35}} \otimes \ket{j_{54},-m_{45}} \, ,
\end{align*}
by attaching the signs $(-1)^{j_{ab}-m_{ab}}$ in \eqref{eq:bell-j} always to the source node, at all links. Then the projected node states are:
\begin{align*}
P_1 \ket{\mathcal{B}_{\gamma_1}(m_{ab})} &= \sum_{i_1} \bar{[i_1]}^{m_{12},m_{13},m_{14},m_{15}} \ket{i_1} \, , \\
P_2 \ket{\mathcal{B}_{\gamma_2}(m_{ab})} &= \sum_{i_2} (-1)^{j_{12}-m_{12}} \bar{[i_2]}^{-m_{12},m_{23},m_{24},m_{25}} \ket{i_2} \, ,  \\
P_3 \ket{\mathcal{B}_{\gamma_3}(m_{ab})} &= \sum_{i_3} \left[ \prod_{b<3} (-1)^{j_{b3}-m_{b3}} \right] \bar{[i_3]}^{-m_{13},-m_{23},m_{34},m_{35}} \ket{i_3} \, ,  \\
P_4 \ket{\mathcal{B}_{\gamma_4}(m_{ab})} &= \sum_{i_4} \left[ \prod_{b<4} (-1)^{j_{b4}-m_{b4}} \right] \bar{[i_4]}^{-m_{14},-m_{24},-m_{34},m_{45}} \ket{i_4} \, ,  \\
P_5 \ket{\mathcal{B}_{\gamma_5}(m_{ab})} &= \sum_{i_5} \left[ \prod_{b<5} (-1)^{j_{b5}-m_{b5}} \right] \bar{[i_5]}^{-m_{15},-m_{25},-m_{35},-m_{45}} \ket{i_5} \, .
\end{align*}
Taking their tensor product and summing over the indices $m_{ab}$, we obtain:
\begin{align}
\ket{\Gamma_5,\mathcal{B}_{\gamma},j_{ab}} =& \sum_{m_{ab}} \left[ \prod_{c<d} (-1)^{j_{cd}-m_{cd}} \right] \sum_{i_a} \left( \bigotimes_{k=1}^5 \ket{i_k} \right) \bar{[i_1]}^{m_{12},m_{13},m_{14},m_{15}} \bar{[i_2]}^{-m_{12},m_{23},m_{24},m_{25}} \nonumber \\
& \quad \times  \bar{[i_3]}^{-m_{13},-m_{23},m_{34},m_{35}} \bar{[i_4]}^{-m_{14},-m_{24},-m_{34},m_{45}} \bar{[i_5]}^{-m_{15},-m_{25},-m_{35},-m_{45}} \nonumber \\ 
=& \sum_{m_{ab}} \sum_{i_a} \left( \bigotimes_{k=1}^5 \ket{i_k} \right) \bar{[i_1]}^{m_{12},m_{13},m_{14},m_{15}} \bar{[i_2]}_{m_{12}}{}^{m_{23},m_{24},m_{25}} \nonumber \\
& \quad \times  \bar{[i_3]}_{m_{13},m_{23}}{}^{m_{34},m_{35}} \bar{[i_4]}_{m_{14},m_{24},m_{34}}{}^{m_{45}} \bar{[i_5]}_{m_{15},m_{25},m_{35},m_{45}} \nonumber \\ 
=& \sum_{i_a} \left( \bigotimes_{k=1}^5 \ket{i_k} \right) \overline{15j(j_{ab},i_a)} \, ,
\end{align}
where the $15j$ symbol is the contraction of the intertwiners determined by the graph $\Gamma_5$:
\be
15j = [i_1]^{m_{12}m_{13}m_{14}m_{15}} [i_2]_{m_{12}}{}^{m_{23}m_{24}m_{25}}[i_3]_{m_{13}m_{23}}{}^{m_{34}m_{35}} [i_4]_{m_{14}m_{24}m_{34}}{}^{m_{45}} [i_5]_{m_{15}m_{25}m_{35}m_{45}} \, ,
\ee
with indices raised and lowered with the $\epsilon$-isomorphism. \par

\section{Hessian of  the $\{15j\}$-symbol action}
We derive explicitly the Hessian of the action for $15j$-symbol. The action for the asymptotic problem of the $15j(j_{ab},\vect{n}_{ab})$ symbol is
\be
S_{(j,\vect{n})}[X] = \sum \limits_{b < a} 2 j_{ab} \, \mathrm{ln} \, \bra{\zeta \vect{n}_{ab}} X_a^{-1} X_{b} \ket{\vect{n}_{ab}} \, ,
\ee
where it encodes a global $SU(2)$ continuous symmetry and a discrete $\pm$ symmetry at each vertex a, given by  $X'_a = \epsilon_a Y X_a$ with $Y \in SU(2)$ and $\epsilon_a = \pm$. 
In order to find the critical points to the action, we need to compute the variation of the action with respect to the $SU(2)$ group elements $X_a$. The variation of the $SU(2)$ group element is simply $\delta X = \tau X$ with the variation of its inverse $\delta X^{-1} = - X^{-1} \tau$, where $\tau_i = \frac{1}{2} i \sigma_i$ is the $su(2)$ algebra element. Therefore, the partial derivative of the action with respect to a $SU(2)$ element $X_d^{j}$ is 
\bea
\frac{\partial S_{(j,\vect{n})}[X]}{\partial X_d^j} = \sum \limits_{b < a} 2 j_{ab} \Bigg\{ \frac{\bra{- \vect{n}_{db}} X_d^{-1} (-\tau_j)  \delta_{ad} X_b \ket{\vect{n}_{bd}}}{\bra{- \vect{n}_{db}} X_d^{-1} X_b \ket{\vect{n}_{bd}}} + \frac{\bra{- \vect{n}_{ad}} X_a^{-1} \tau_j \delta_{bd} X_d \ket{\vect{n}_{da}}  }{\bra{- \vect{n}_{ad}} X_a^{-1} X_d \ket{\vect{n}_{da}} } \Bigg\} \\
= \sum \limits_{b < d} 2 j_{db}  \frac{\bra{- \vect{n}_{db}} X_d^{-1} (-\tau_j) X_b \ket{\vect{n}_{bd}}}{\bra{- \vect{n}_{db}} X_d^{-1} X_b \ket{\vect{n}_{bd}}} + \sum \limits_{a > d} 2 j_{ad}  \frac{\bra{- \vect{n}_{ad}} X_a^{-1} \tau_j  X_d \ket{\vect{n}_{da}}  }{\bra{- \vect{n}_{ad}} X_a^{-1} X_d \ket{\vect{n}_{da}} } \, ,
\eea
where $\ket{-\vect{n}_{ab}}$ is obtained by the action of the antilinear map $\zeta$ on coherent states, which takes $\vect{n}$ to $-\vect{n}$. The stationarity of the action $\delta S_{(j,\vect{n})}[X] = 0$ leads to a set of complex vector equations: 
\be
\sum \limits_{b \neq a} j_{ab} \vect{v}_{ab} = 0, \,\,\, \vect{v}_{ab} = - \vect{v}_{ba} \, ,
\ee
where the vector $\vect{v}_{ab}$ is defined as
\be
\label{eq:vector}
\vect{v}_{ab} = \frac{\bra{-\vect{n}_{ab}} X_a^{-1} \boldsymbol{\sigma} X_b \ket{\vect{n}_{ba}}}{\bra{-\vect{n}_{ab}} X_a^{-1} X_b \ket{\vect{n}_{ba}}}
\ee
and the minus sign of the second term in the variation of the action can be taken in a single expression by knowing that the epsilon inner product $\epsilon(T \alpha, \beta) = - \epsilon(\alpha, T \beta)$ for algebra element $T$ and  $\epsilon(g \alpha, \beta) = \epsilon(\alpha, g^{-1} \beta)$ for group element $g$. As the action of the group element on coherent states $\ket{\vect{n}_{ab}}$ produces new set of coherent states $\ket{\vect{n'}_{ab}}$, $X_a \ket{\vect{n}_{ab}} = \ket{\vect{n'}_{ab}}$, we can simplify the vector $\vect{v}_{ab}$ expression:
\be
\vect{v}_{ab} = \frac{\bra{-\vect{n'}_{ab}} \boldsymbol{\sigma} \ket{ \vect{n'}_{ba}}}{\bra{-\vect{n'}_{ab}} \vect{n'}_{ba}\rangle } = \frac{(\vect{n'}_{ba} - \vect{n'}_{ab}) - i (- \vect{n'}_{ab} \times \vect{n'}_{ba})}{1 - \vect{n'}_{ab} \vect{n'}_{ba}} = -\vect{n'}_{ab} \, ,
\ee
where we used the scalar product of coherent states and the projector $P_n = \ket{\vect{n}} \bra{\vect{n}} = \frac{1}{2}(\mathbb{I} + \boldsymbol{\sigma} \cdot \vect{n})$. It is clear that the stationary methods can be extended for real part of the function so that the real part of the action is maximized by setting the imaginary part to zero, $\vect{n'}_{ab}$ and $\vect{n'}_{ba}$ should be anti-parallel,
$X_a \vect{n}_{ab} = - X_b \vect{n}_{ba}$. Therefore we have ten equations
\bea
\label{critical1}
X_a \vect{n}_{ab} = - X_b \vect{n}_{ba}
\eea 
for maximizing the action and five equations for the stationarity of the action 
\bea
\label{critical2}
\sum \limits_{b \neq a} j_{ab} \vect{n}_{ab} = 0 \, .
\eea
The solutions to the critical equations \eqref{critical1}-\eqref{critical2} contain an interpretation in terms of the $BF$ theory. They can be considered as the solutions of a four-dimensional $BF$ theory with group $SU(2)$ discretized on a $4$-simplex. Therefore, the solutions can be parametrized by the $X_a$ and $\vect{b}_{ab}=j_{ab} X_a \vect{n}_{ab}$. These are the discrete connection and B-field variables, respectively. The critical equations are now expressed in terms of the $\vect{b}_{ab}$:
\bea
\sum \limits_{b:b \neq a} \vect{b}_{ab} = 0 , \,\,\,\,\, \vect{b}_{ab} = - \vect{b}_{ba} \, ,
\eea
which determines a vector geometry by twenty three-dimensional vectors $\vect{b}_{ab}$. The $X_a$ variables are a discrete version of the connection on a $BF$ theory. This is due to the gluing of two $4$-simplexes followed by the identification of the $\vect{n}_{ab}$ variables on a common tetrahedron. In terms of the vector geometry ($\vect{b}_{ab}$ variables), this implies that the gluing takes place after the action of the corresponding $X_a$ for the tetrahedron. \par
As we have the critical points to the action, we can evaluate the Hessian of the action. The Hessian is the second derivative of the action with respect to the group element:
\bea
H_{cd}^{ij} \equiv \frac{\partial ^2 S_{(j,\vect{n})}[X]}{\partial X_c^i \partial X_d^j} = \sum \limits_{b < d} 2 j_{db} \Bigg\{ \frac{\bra{- \vect{n}_{cb}} X_c^{-1} (\tau_i \tau_j) \delta_{cd} X_b \ket{\vect{n}_{bc}}}{\bra{- \vect{n}_{cb}} X_c^{-1} X_b \ket{\vect{n}_{bc}}} + \frac{\bra{- \vect{n}_{dc}} X_d^{-1} (-\tau_j \tau_i) \delta_{bc} X_c \ket{\vect{n}_{cd}}}{\bra{- \vect{n}_{dc}} X_d^{-1} X_c \ket{\vect{n}_{cd}}} \nonumber \\
- \frac{\bra{- \vect{n}_{db}} X_d^{-1} (- \tau_j) X_b \ket{ \vect{n}_{bd}}}{\bra{-\vect{n}_{db}} X_d^{-1} X_b \ket{\vect{n}_{bd}}^2} \Big( \bra{-\vect{n}_{cb}} X_c^{-1} (-\tau_i) \delta_{cd} X_b \ket{\vect{n}_{bc}} + \bra{-\vect{n}_{dc}} X_d^{-1} \tau_i \delta_{bc} X_c \ket{\vect{n}_{cd}} \Big) \Bigg\} \nonumber \\
+  \sum \limits_{a > d} 2 j_{ad} \Bigg\{ \frac{\bra{- \vect{n}_{cd}} X_c^{-1} (-\tau_i \tau_j) \delta_{ca} X_d \ket{\vect{n}_{dc}}}{\bra{- \vect{n}_{cd}} X_c^{-1} X_d \ket{\vect{n}_{dc}}} + \frac{\bra{- \vect{n}_{ac}} X_a^{-1} (\tau_j \tau_i) \delta_{cd} X_c \ket{\vect{n}_{ca}}}{\bra{- \vect{n}_{ac}} X_a^{-1} X_c \ket{\vect{n}_{cd}}} \nonumber \\
- \frac{\bra{- \vect{n}_{ad}} X_a^{-1} (\tau_j) X_d \ket{ \vect{n}_{da}}}{\bra{-\vect{n}_{ad}} X_a^{-1} X_d \ket{\vect{n}_{da}}^2} \Big( \bra{-\vect{n}_{cd}} X_c^{-1} (-\tau_i) \delta_{ac} X_d \ket{\vect{n}_{dc}} + \bra{-\vect{n}_{ac}} X_a^{-1} \tau_i \delta_{cd} X_c \ket{\vect{n}_{ca}} \Big) \Bigg\} \nonumber
\eea
for the case $c < d$ and $c > d$ with using the definition of $\vect{v}_{ab}$ we have:
\begin{align}
\label{Hessian1}
c < d : -\frac{1}{2} j_{cd} \Bigg[\frac{\bra{- \vect{n}_{dc}} X_d^{-1} (-\delta_{ij} - i \epsilon_{ijk} \sigma_k) X_c \ket{\vect{n}_{cd}}}{\bra{- \vect{n}_{dc}} X_d^{-1} X_c \ket{\vect{n}_{cd}}} + v_{dc}^j v_{dc}^i \Bigg] 
 = - \frac{1}{2} j_{cd} \big(-\delta_{ij} - i \epsilon_{ijk} v_{cd}^k + v_{cd}^i v_{cd}^j \big) \nonumber \\
c > d : -\frac{1}{2} j_{cd} \Bigg[\frac{\bra{- \vect{n}_{cd}} X_c^{-1} (-\delta_{ij} - i \epsilon_{ijk} \sigma_k) X_d \ket{\vect{n}_{dc}}}{\bra{- \vect{n}_{cd}} X_c^{-1} X_d \ket{\vect{n}_{dc}}} + v_{cd}^j v_{cd}^i \Bigg]
= - \frac{1}{2} j_{cd} \big(-\delta_{ij} - i \epsilon_{ijk} v_{cd}^k + v_{cd}^i v_{cd}^j \big)  \, ,
\end{align}
and for $c = d$:
\begin{align}
c = d & : -\frac{1}{4} \Bigg\{ \sum \limits_{b < c} 2 j_{bc} \bigg(\frac{\bra{- \vect{n}_{bc}} X_c^{-1} (\delta_{ij} + i \epsilon_{ijk} \sigma_k) X_b \ket{\vect{n}_{bc}}}{\bra{- \vect{n}_{cb} X_c^{-1} X_b \ket{\vect{n}_{bc}}}} - v_{cb}^j v_{cb}^i \bigg) \nonumber \\
&+ \sum \limits_{a > c} 2 j_{ac} \bigg( \frac{\bra{- \vect{n}_{ac}} X_a^{-1} (\delta_{ij} + i \epsilon_{ijk} \sigma_k) X_c \ket{\vect{n}_{ca}}} {\bra{-\vect{n}_{ac}} X_a^{-1} X_c \ket{\vect{n}_{ca}}}  - v_{ac}^j v_{ac}^i \bigg) \Bigg\} \nonumber \\
& = - \frac{1}{4} \sum \limits_{b \neq c} 2 j_{bc} (\delta_{ij} + i \epsilon_{ijk} v_{cb}^k - v_{cb}^i v_{cb}^j) \, ,
\label{Hessian2}
\end{align}
where we have used the identity $\sigma_i \sigma_j  = \delta_{ij} + i \epsilon_{ijk} \sigma_k$. The Hessian of the action, as a second derivative test of a function, can be used to express the asymptotic expansion of the integral over the group elements $X$, which corresponds to 15j-symbol proportional to $e^{\lambda S}$ given by the boundary data:
\be
\{15j\}(\lambda j,n) \sim \bigg(\frac{2 \pi}{\lambda}\bigg)^6 \frac{1}{\sqrt{\mathrm{det} \, H}} \, e^{\lambda S(X)} \, ,
\ee
where the action $S$ for the $15j$-symbol is evaluated at the critical points derived above. In order to apply the method of extended stationary phase we need to ensure that the stationary points are isolated. This is accomplished by fixing $SU(2)$ elements as: $X'_a = (X_5)^{-1}X_a$ for $a=\{1,\cdots,4 \}$. The ``gauge fixed'' integral formulas then have isolated critical points related only by the discrete symmetries and can now be evaluated using extended stationary phase. After gauge fixing and deriving the expression of the Hessian $H_{cd}^{ij}$ for a given boundary data  $\{j_{ab},\vect{n}_{ab}\}$, we can perform the explicit calculation of the Hessian matrix $H$:
\bea
H = \begin{bmatrix}
     H_{11} & H_{12} & \dots  & H_{14} \\
     H_{21} & H_{22} & \dots  & H_{24} \\
    \vdots & \vdots & \vdots & \vdots \\
    H_{41} & H_{42} & \dots  & H_{44} 
\end{bmatrix} \, .
\eea

\section{Vector geometry for equal spins $j_{ab}=j$}
\label{appendix:Vector geometry for equal spins}

Let us illustrate the structure of the space of vector geometries in the concrete case of equiareal tetrahedra. It is convenient to introduce the unit vectors $\vect{w}_{ab}:=X_a \vect{n}_{ab}$. We will determine the dimensionality of the space of solutions to the critical point equations and discuss the relevant subspaces. From the condition $\vect{w}_{ab}=-\vect{w}_{ba}$, only ten of the twenty vectors $\vect{w}_{ab}$ can be independent---one per link. Consider the node $1$. There are four vectors $\vect{w}_{1b}$ at the node. We can use the symmetry under global rotations to fix:
\be
\vect{w}_{12} = \hat{\vect{z}} \, , \qquad  \vect{w}_{13}=(\sin \theta, 0, \cos \theta) \, ,
\ee
so that the two first vectors are described by a single parameter.
\begin{itemize}
\item For $\theta=0$, we have $\vect{w}_{12}=\vect{w}_{13}=\hat{\vect{z}}$, and the closure relation gives $\vect{w}_{14}=\vect{w}_{15}=-\hat{\vect{z}}$. This is a degenerate case, where the normals form a linear object. This subspace contains a single solution.
\item For $\theta=\pi$, $\vect{w}_{13}=-\vect{w}_{12}=-\hat{\vect{z}}$, and the closure relation gives $\vect{w}_{15}=-\vect{w}_{14}$, where $\vect{w}_{14}$ can be freely chosen. The symmetry under global rotations can be used to force one of the components of $\vect{w}_{14}$ to vanish, leaving one free parameter. Hence, this subspace is one-dimensional. If $\vect{w}_{14}=\pm \hat{\vect{z}}$, then the geometry is the same as for $\theta=0$. The normals form a planar object, and we have again a degenerate geometry.
\item Now take $\theta\neq 0,\pi$. From the closure relation $\sum_{b\neq 1} \vect{w}_{1b}=0$, it follows that the partial sum $\vect{w}_{12} +  \vect{w}_{13} + \vect{w}_{14}$ must be a unit vector, equal to $-\vect{w}_{15}$. The possible choices of $\vect{w}_{14}$ form a unit sphere centered at $0+\vect{w}_{12}+\vect{w}_{13}$, which intersects the unit sphere centered at the origin $0$ at a one-dimensional loop. Therefore, we have one free parameter associated with $\vect{w}_{14}$, and the space of solutions to the closure relation, up to global rotations, is two-dimensional. If one takes $\vect{w}_{14}=-\vect{w}_{13}$ or $-\vect{w}_{12}$, then the normals form a parallelogram, and we have a degenerate geometry. For any other choice, the normals form a nondegenerate geometry of nonzero volume, describing a unique tetrahedron.
\end{itemize}
In short, the space of solutions of the closure relation at a node, up to rotations, is two-dimensional, and the subset of degenerate solutions forms a lower dimensional subspace.

Consider now the node $2$. We have $\vect{w}_{21}=-\vect{w}_{12}$. The vector $\vect{w}_{23}$ is completely free, requiring two free parameters for its description. As before, we have a single extra parameter for the description of $\vect{w}_{24}$, in order for the closure relation to admit a solution. The last vector $\vect{w}_{25}$ is then fixed by the closure relation. We have three additional parameters associated with the second node.

For the node $3$, the vectors $\vect{w}_{31}$ and $\vect{w}_{32}$ are fixed by the previous choices of normals for the nodes $1$ and $2$. We have again one free parameter for the possible choices of $\vect{w}_{34}$. However, the same is true for $\vect{w}_{43}=-\vect{w}_{34}$ at the node $4$, and the intersection of two loops on the sphere is in general formed by isolated points, and we do not have a new degree of freedom associated with these nodes. All remaining vectors at the nodes $3$ and $4$ are then fixed by the closure relations. The closure relation at the node $5$ imposes an additional condition, but this is automatically satisfied when the closure relations at the nodes $1$ to $4$ and the link conditions are satisfied.

We conclude that the space of solutions to the critical point conditions on the pentagram with all spins equal, $j_\ell=j$, is characterized by five parameters, and the subset of degenerate geometries forms a lower-dimensional subset at the boundary of the parameter space.

\section{Shape-matched configurations for a 4-simplex in $\mathbb{R}^3$}
\label{appendix:Shape-matching configurations for a 4-simplex}

The boundary data of a 4-simplex in $\mathbb{R}^3$ is equipped with a metric of signature $0+++$ and has the same metric geometry as a linear immersion of the simplex into $\mathbb{R}^3$. For a single solution set $X_a$ to the critical point equations, the gluing map $g_{ab}$ can be either an identity or $\pi$ rotation with which the boundary data coincides with the $3d$ Euclidean geometry. In other words, the shape-matched normals in $3d$ space should form the boundary of a 4-simplex via geometric way. 
We will construct the boundary of a 4-simplex in $\mathbb{R}^3$ out of local modifications to a 3-manifold triangulation, where a collection of five tetrahedra whose twenty faces are glued together in ten pairs. There are four such modifications knowns as Pachner moves: \textbf{1-4} move is replacing a single tetrahedron with four distinct tetrahedra meeting at a common internal vertex. \textbf{2-3} move is taking two distinct tetrahedra joined along a common face with three distinct tetrahedra joined along a common edge. The remaining moves, 3-2 and 4-1 moves, are just inverse to the 2-3 and 1-4 moves. These moves do not change the topology of the triangulation at all. We will consider only 1-4 and 2-3 moves since their critical point equations or the orientation conditions for the relevant vector geometry are exactly the same as in 4-1 and 3-2, respectively. \par
\begin{figure}[htbp]
\begin{center}
 \includegraphics[scale=1.25]{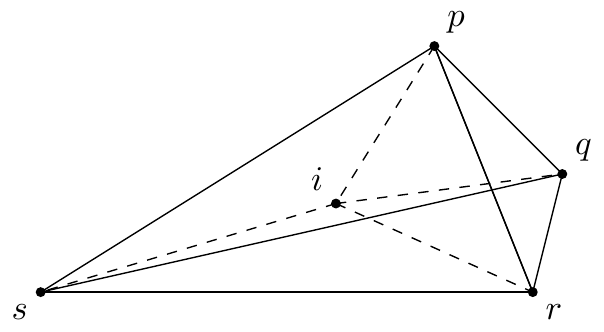}
\end{center} 
\caption{Description of the 1-4 move used to construct five glued tetrahedra by placing a fifth vertex $i$, origin of the coordinates, in the center point of the tetrahedron $pqrs$.}
\end{figure}

\begin{figure}[htbp]
\begin{center}
 \includegraphics[scale=1.2]{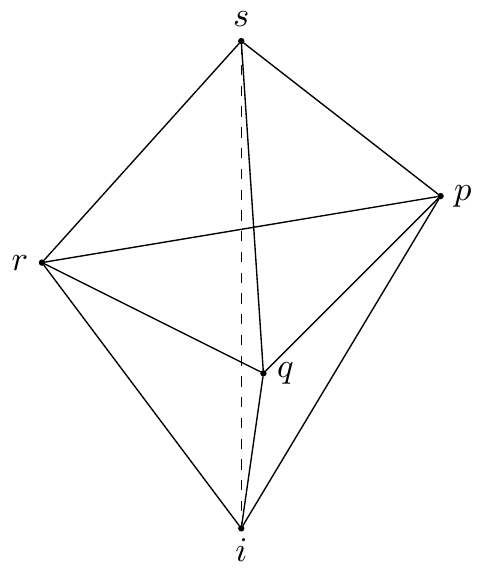}
\end{center} 
\caption{Pachner move 2-3.  Two tetrahedra glued back to back are transformed into three tetrahedra with the inclusion of a new edge from the vertex $s$ to the vertex $i$.}
\label{fig:2-3}
\end{figure}

Consider a regular tetrahedron $\tau_f$ with side length 2 by having four position vectors of four vertices where the regular tetrahedron is centered at the origin of the coordinates:
\bea
\vect{p}^T = (1,0, - \frac{1}{\sqrt{2}}), \,\,\, \vect{q}^T = (-1,0, - \frac{1}{\sqrt{2}}), \,\,\, \vect{r}^T = (0,1, \frac{1}{\sqrt{2}}), \,\,\, \vect{s}^T = (0, -1, \frac{1}{\sqrt{2}}) \, .
\eea
We can construct the edge vectors $\vect{e}_k$ in order to fully determine the regular tetrahedron:
\bea
\vect{e}_1 = \vect{p}-\vect{s}, \,\,\, \vect{e}_2 = \vect{q}-\vect{s}, \,\,\, \vect{e}_3 = \vect{r}-\vect{s} \, .
\eea
One can obtain the outward normals of the regular tetrahedron ($\vect{E}= A^{1-4}_{\mathrm{out}} \, \vect{n}, A^{1-4}_{\mathrm{out}}= \sqrt{3}$) from the ``electric field'' on each face of the tetrahedron:
\bea
\label{electric-field}
\vect{E}_1 = \frac{1}{2} (\vect{e}_2 \times \vect{e}_3), \,\,\, \vect{E}_2 = \frac{1}{2} (\vect{e}_3 \times \vect{e}_1), \,\,\, \vect{E}_3 = \frac{1}{2}  (\vect{e}_1 \times \vect{e}_2)
\eea
with satisfying the closure condition on the regular tetrahedron $\vect{E}_4 = -(\vect{E}_1 + \vect{E}_2 + \vect{E}_3)$ and the normals related to the exterior faces of $\tau_f$ are then:
\bea
\label{exterior_normals_sm1}
\vect{n}_{cf} = (1/A^{1-4}_{\mathrm{out}}) \, \vect{E}_3, \,\, \vect{n}_{af} = (1/A^{1-4}_{\mathrm{out}}) \, \vect{E}_1, \,\, \vect{n}_{bf} = (1/A^{1-4}_{\mathrm{out}}) \, \vect{E}_2, \,\,  \vect{n}_{df} = -(\vect{n}_{cf} + \vect{n}_{af} + \vect{n}_{bf}) \, ,
\eea
where the labelings $a,b,c,d$ correspond to the four interior tetrahedra $\tau_a, \tau_b, \tau_c, \tau_d$, respectively. The normals of the interior faces with area $A^{1-4}_{\mathrm{in}}=1/\sqrt{2}$ belonging to the four interior tetrahedra are:
\bea
\tau_b : \vect{n}_{cb} = \frac{1}{2 A^{1-4}_{\mathrm{in}}} (\vect{d}-\vect{f}) \times (\vect{a}-\vect{f}), \,\,\, \vect{n}_{db}=\frac{1}{2 A^{1-4}_{\mathrm{in}}} (\vect{a}-\vect{f}) \times (\vect{c}-\vect{f}), \nonumber \\ \vect{n}_{ab}=\frac{1}{2 A^{1-4}_{\mathrm{in}}} (\vect{c}-\vect{f}) \times (\vect{d}-\vect{f}), \,\,\, \vect{n}_{fb}=\vect{n}_{bf} \nonumber
\eea
\bea
\tau_d: \vect{n}_{ad}=\frac{1}{2 A^{1-4}_{\mathrm{in}}} (\vect{b}-\vect{f}) \times (\vect{c}-\vect{f}), \,\,\, \vect{n}_{cd}=\frac{1}{2 A^{1-4}_{\mathrm{in}}} (\vect{a}-\vect{f}) \times (\vect{b}-\vect{f}), \nonumber \\ \vect{n}_{bd}=- \vect{n}_{db}, \,\,\, \vect{n}_{fd} = \vect{n}_{df} \nonumber
\eea
\bea
\tau_a : \vect{n}_{ca}=\frac{1}{2 A^{1-4}_{\mathrm{in}}} (\vect{b}-\vect{f}) \times (\vect{d}-\vect{f}), \,\,\, \vect{n}_{fa}=\vect{n}_{af}, \,\,\, \vect{n}_{da}=-\vect{n}_{ad}, \,\,\, \vect{n}_{ba} = -\vect{n}_{ab}  \nonumber
\eea
\bea
\tau_c:  \vect{n}_{fc} = \vect{n}_{cf}, \, \vect{n}_{ac} = - \vect{n}_{ca}, \,\,\, \vect{n}_{dc} = - \vect{n}_{cd}, \,\,\, \vect{n}_{bc} = - \vect{n}_{cb} \, ,
\eea 
where there are in total 4 aligned and 6 back to back conditions on the normals. These conditions can be applied to the configurations other than shape-matched in the 1-4 vector geometry. As the normals are obtained on a shape-matched configuration for a 4-simplex, we can compute $|\mathrm{det} H|^{-1/2}$ for fixed spins $j_{\mathrm{out}}=2, j_{\mathrm{in}}=2/\sqrt{6}$, which is related to the ratio of the areas ($A_{\mathrm{out}},A_{\mathrm{in}})$ of the exterior and interior faces, on this boundary data:
\bea
|\mathrm{det} H_0|^{1/2}_{1-4} = 0.204947 \, .
\eea 
Now we can consider 2-3 move to construct a 4-simplex in $\mathbb{R}^3$ which shown in Fig.~\ref{fig:2-3}.

Consider two regular tetrahedra glued back to back with side length of 2 by having five position vectors of five vertices where the regular tetrahedron at the top is centered at the origin of the coordinates:
\bea
\vect{p}^T = (1,0, - \frac{1}{\sqrt{2}}), \, \vect{q}^T = (-1,0, - \frac{1}{\sqrt{2}}), \, \vect{r}^T = (0,1, \frac{1}{\sqrt{2}}), \, \vect{s}^T = (0, -1, \frac{1}{\sqrt{2}}), \, \vect{i}^T=(0,\frac{5}{3}, - \frac{5}{3\sqrt{2}}) \, . \nonumber
\eea
The normals of the two regular tetrahedra $\tau_f$ and $\tau_d$  with exterior and interior areas $A^{2-3}_{\mathrm{out}}=\sqrt{3},A^{2-3}_{\mathrm{in}}=(4/3)\sqrt{2}$ are then respectively as follows
\bea
\vect{n}_{cf} = (1/A^{2-3}_{\mathrm{out}}) \, \vect{E}_3, \,\, \vect{n}_{af} = (1/A^{2-3}_{\mathrm{out}}) \, \vect{E}_1, \,\, \vect{n}_{bf} = (1/A^{2-3}_{\mathrm{out}}) \, \vect{E}_2, \,\,  \vect{n}_{df} = -(\vect{n}_{cf} + \vect{n}_{af} + \vect{n}_{bf}) \nonumber
\eea
\bea
\vect{n}_{cd} = (1/A^{2-3}_{\mathrm{out}}) \, \vect{F}_3, \,\, \vect{n}_{ad} = (1/A^{2-3}_{\mathrm{out}}) \, \vect{F}_1, \,\, \vect{n}_{bd} = (1/A^{2-3}_{\mathrm{out}}) \, \vect{F}_2, \,\,  \vect{n}_{fd} = -\vect{n}_{df} \, ,
\eea
where the vectors $\vect{E}_k$ are given in \eqref{electric-field} and $\vect{F}_k$ have the same expression as in the $\vect{E}_k$ via replacing the vertex $s$ by the vertex $i$. The sets of the normals belonging to three interior tetrahedra $\tau_c$, $\tau_a$ and $\tau_b$ are as follows
\bea
\tau_c : \vect{n}_{bc} = \frac{1}{2 A^{2-3}_{\mathrm{in}}} (\vect{a}-\vect{f}) \times (\vect{d}-\vect{f}), \, \vect{n}_{ac}=\frac{1}{2 A^{2-3}_{\mathrm{in}}} (\vect{d}-\vect{f}) \times (\vect{b}-\vect{f}), \, \vect{n}_{fc}= \vect{n}_{cf}, \, \vect{n}_{dc}=\vect{n}_{cd} \nonumber
\eea
\bea
\tau_a : \vect{n}_{ba} = \frac{1}{2 A^{2-3}_{\mathrm{in}}} (\vect{d}-\vect{f}) \times (\vect{c}-\vect{f}), \, \vect{n}_{da}= \vect{n}_{ad} , \, \vect{n}_{fa}= \vect{n}_{af}, \, \vect{n}_{ca}=-\vect{n}_{ac} \nonumber
\eea
\bea
\tau_b : \vect{n}_{fb} = \vect{n}_{bf}, \, \vect{n}_{db}= \vect{n}_{bd} , \, \vect{n}_{ab}= -\vect{n}_{ba}, \, \vect{n}_{cb}=-\vect{n}_{bc}  \, .
\eea
In the 2-3 vector geometry, there are in total 6 aligned and 4 back to back conditions on the normals and the $|\mathrm{det} H|^{-1/2}$ for fixed spins $j_{\mathrm{out}}=2,j_{\mathrm{in}}=j_{\mathrm{out}} \, (4/3) \sqrt{2/3}$ on this boundary data is
\bea
\label{eq:hessian_2-3}
|\mathrm{det} H_0|^{1/2}_{2-3} = 0.262621 \, .
\eea 


\bibliography{entangledpolyhedra-arxiv-v1}

\end{document}